\documentclass[11pt]{article}
\usepackage{longtable}
\usepackage{ae,aecompl,pslatex}
\usepackage[dvips]{graphicx}
\usepackage{enumerate,color}
\usepackage{fullpage}
\usepackage{lineno}
\usepackage[table]{xcolor}
\definecolor{lightgray}{gray}{0.9}
\usepackage{float}

\usepackage{hyperref}
\hypersetup{
    colorlinks=true,
    citecolor=green,  
}

\makeatletter
\DeclareRobustCommand{\linkedyearcite}[3]{%
    #1 \hyperlink{cite.#2}{#3}%
}
\makeatother
\usepackage{rotating}
\usepackage{placeins}
\usepackage{psfrag,epsfig,graphicx,color,array,dcolumn,tabularx,delarray,longtable,indentfirst,amsmath,amsfonts,amssymb,amsthm,eucal,bbm}
\theoremstyle{bkaexa} 
\usepackage{caption}
\usepackage{setspace}
\theoremstyle{bkaexa} 
\newtheorem{Rem}{Remark}
\theoremstyle{bkathm} 

\theoremstyle{bkathm} 
\newtheorem{Thm}{Theorem}
\theoremstyle{bkathm} 
\newtheorem{Cor}{Corollary}
\theoremstyle{bkathm} 
\newtheorem{Lem}{Lemma}
\theoremstyle{definition}

\begin{document}
\setstretch{1.5}
\title{Categorical distance correlation under general encodings and its application to high-dimensional feature screening}
\author{\normalsize Qingyang Zhang\\
\normalsize Department of Mathematical Sciences, University of Arkansas, Fayetteville, AR 72701, USA\\
\normalsize Email: qz008@uark.edu
}
\date{}
\maketitle

\begin{abstract}
In this paper, we extend distance correlation to categorical data with general encodings, such as one-hot encoding for nominal variables and semicircle encoding for ordinal variables. Unlike existing methods, our approach leverages the spacing information between categories, which enhances the performance of distance correlation. Two estimates including the maximum likelihood estimate and a bias-corrected estimate are given, together with their limiting distributions under the null and alternative hypotheses. Furthermore, we establish the sure screening property for high-dimensional categorical data under mild conditions. We conduct a simulation study to compare the performance of different encodings, and illustrate their practical utility using the 2018 General Social Survey data.
\end{abstract}


\section{Introduction}
Categorical data is fundamental across a vast range of scientific disciplines, from social sciences to biology, medicine, and marketing. For instance, in medicine, it is routinely employed to track patient recovery stages, and classify treatment efficacy. Similarly, in marketing, companies often leverage categorical data to group customers by preferences or brand loyalty, informing targeted strategies. Biological researchers frequently utilize it to classify species, observe genetic traits (e.g., presence or absence of a specific gene), and categorize diverse ecosystem types. Pearson's chi-squared test is one of the most well established and widely applied statistical procedures for analyzing categorical data. However, it may suffer from low power when applied to relatively large and sparse contingency tables, as the underlying normality assumption of the test may be violated. To overcome this limitation, two permutation tests have been recently developed, including the distance covariance permutation test\linkedyearcite{}{Zhang2019}{(Zhang, 2019)} and the $U$-statistic permutation test \linkedyearcite{}{USP}{(Berrett \& Samworth, 2021};\linkedyearcite{}{USPAOS}{Berrett et al., 2021)}. These tests have shown greater power than Pearson's chi-squared test under limited sample sizes (e.g., see Figure 2 in\linkedyearcite{}{Zhang2019}{Zhang, 2019}, and Figure 3 in\linkedyearcite{}{USP} {Berrett \& Samworth, 2021}). Both tests utilize the following simple functional but different estimates (Zhang used the maximum likelihood estimate while Berrett \& Samworth used the minimum variance unbiased estimate)
\begin{equation} \label{e1}
\eta (X, Y) = \sum_{i=1}^{I}\sum_{j=1}^{J}(\pi_{ij} - \pi_{i+}\pi_{+j})^2,
\end{equation}
where $X$ and $Y$ are two categorical variables with joint probabilities $\{\pi_{ij}\}_{1\leq i\leq I, 1\leq j\leq J}$, and marginal probabilities $\{\pi_{i+}\}_{1\leq i\leq I}$ and $\{\pi_{+j}\}_{1\leq j\leq J}$, respectively. It is straightforward to verify that the $\eta$ characterizes independence, i.e., $\eta (X, Y) = 0$ if and only if $X$ and $Y$ are independent. However, unlike Pearson's chi-squared divergence, $\eta$ is essentially an $L_2$ divergence which is always well-defined even when some marginal probabilities are close to zero. Notably, $\eta$ is equivalent to (up to a constant) the following squared distance covariance
\begin{equation} \label{e2}
\mathrm{dCov}^{2}(X, Y) = E(\|X_{1} - X_{2}\|\|Y_{1} - Y_{2}\|) + E(\|X_{1} - X_{2}\|)E(\|Y_{1} - Y_{2}\|) - 2E(\|X_{1} - X_{2}\|\|Y_{1} - Y_{3}\|),
\end{equation} 
where $(X_{1}, Y_{1})$, $(X_{2}, Y_{2})$ and $(X_{3}, Y_{3})$ are three independent copies of $(X, Y)$, and $\|\cdot\|$ is the Euclidean distance. The expression in Equation \ref{e2} was derived from the weighted $L_{2}$ distance between the joint characteristic function and the product of marginal characteristic functions, under certain choice of weights\linkedyearcite{}{DC}{(Sz\'{e}kely et al., 2007)}.

Building upon the work of\linkedyearcite{}{Zhang2019}{Zhang (2019)} and\linkedyearcite{}{USP}{Berrett \& Samworth (2021)},\linkedyearcite{}{Zhang2025}{Zhang (2025)} further established some key properties of $\eta$, including its $B$-robustness under both fixed and diverging dimensions, and screening consistency under certain conditions. To enable analytical p-value calculations,\linkedyearcite{}{Zhang2025}{Zhang (2025)} also derived an approximate null distribution for the unbiased estimate. The asymptotic null distribution of the maximum likelihood estimate was derived by\linkedyearcite{}{CP2024}{Castro-Prado et al. (2024)}. Despite its simplicity and nice properties, a major limitation of the functional $\eta$ is that it is based on one-hot encoding, thus only suitable for nominal variables. For ordinal variables, $\eta$ fails to incorporate the spacings between categories, which may result in loss of information and less statistical power. To address this limitation, in this paper, we extend the functional $\eta$ from one-hot encoding to general encodings, making it suitable for all categorical variables. The asymptotic distributions of two sample estimates are provided. Furthermore, we consider the application to high-dimensional feature screening, and establish the sure screening property under mild conditions. 

Our framework expands the applicability of the categorical distance correlation to a much broader range of scenarios, and can be readily applied to many real datasets that contains different types of categorical variables. For instance, in the General Social Survey (GSS) data, the majority of the variables are categorical, which include both nominal variables (gender, race, marital status etc) and ordinal variables (education level, socioeconomic status etc). Our extended measures and screening procedure can be used to explore the complex relationships between these variables, while incorporating user-defined spacings between categories for each variable.   

The remainder of this paper is structured as follows: In Section \ref{sec2}, we first review some commonly used encodings for categorical data and then derive the categorical distance correlation under general encodings. We then study the properties of two sample estimates: the maximum likelihood estimate and a bias-corrected estimate. In particular, we establish their asymptotic distributions under the null and alternative hypotheses, and their screening consistency under high dimensions. Section \ref{sec3} compares different encodings using simulated data, providing some empirical recommendations regarding the choice of encodings. In Section \ref{sec4}, we apply the proposed method to the 2018 GSS data to identify categorical features that are most relevant to people's subjective socioeconomic classifications. Section \ref{sec5} discusses the paper with some future perspectives.

\section{Methods}\label{sec2}
\subsection{Different encodings for categorical data}\label{sec21}
We begin by reviewing some common encodings for categorical data. Let $X$ be a random variable with $I$ categories, encoded as $\{\mathbf{u}_{1}, ..., \mathbf{u}_{I}\}$. If $X$ is nominal, i.e., the categories have no natural order such as race and gender, one-hot encoding is a common choice. In this encoding, $\mathbf{u}_{i}$ is a vector of size $I$ with the $i$th element being 1 and others 0. For example, for $I=3$, the one-hot encodings would be $\mathbf{u}_{1} = (1, 0, 0)$, $\mathbf{u}_{2} = (0, 1, 0)$, $\mathbf{u}_{3} = (0, 0, 1)$. It is worth noting that one-hot encoding results in an $I$-dimensional representation, and the Euclidean distance between any two distinct categories is $\sqrt{2}$. 

If $X$ is ordinal, i.e., the categories are naturally ordered such as education level, two well-established encodings are ordinal encoding and semicircle encoding\linkedyearcite{}{semicircle}{(Tian et al., 2024)}. In ordinal encoding, each category is assigned an integer value, e.g., the equally spaced ordinal encodings are $\mathbf{u}_{i} = i$, $i = 1, ..., I$. For example, when $I=3$, we have $\mathbf{u}_{1} = 1$, $\mathbf{u}_{2} = 2$ and $\mathbf{u}_{3} = 3$. Unlike one-hot encoding, ordinal encoding has dimension 1, and the spacings are additive, i.e., for any three ordered categories, $i_{1}< i_{2} < i_{3}$, we have $\| \mathbf{u}_{i_{3}} - \mathbf{u}_{i_{1}} \| = \| \mathbf{u}_{i_{2}} - \mathbf{u}_{i_{1}} \| + \| \mathbf{u}_{i_{3}} - \mathbf{u}_{i_{2}} \|$. In semicircle encoding, each category is mapped to a point on the semicircle, e.g., the equally spaced semicircle encodings are
\begin{equation*}
\mathbf{u}_{i} = \left[\cos \frac{(i-1)\pi}{I-1},~ \sin \frac{(i-1)\pi}{I-1} \right], ~i=1, ..., I.
\end{equation*}
For instance, when $I=4$, we have $\mathbf{u}_{1} = (1, 0)$, $\mathbf{u}_{2} = (1/2, \sqrt{3}/2)$, $\mathbf{u}_{3} = (-1/2, \sqrt{3}/2)$, $\mathbf{u}_{4} = (-1, 0)$. It is noteworthy that semicircle encoding has dimension 2, and the spacings satisfy the triangle inequality, i.e., for any three ordered categories, $i_{1}< i_{2} < i_{3}$, we have $\| \mathbf{u}_{i_{3}} - \mathbf{u}_{i_{1}} \| < \| \mathbf{u}_{i_{2}} - \mathbf{u}_{i_{1}} \| + \| \mathbf{u}_{i_{3}} - \mathbf{u}_{i_{2}} \|$.

The encoding of categorical data should reflect the relative spacings between categories. Consider a variable $X\in \{\text{mild,~moderate,~severe}\}$ representing disease severity. If the moderate category is medically perceived as closer to mild than to severe, this understanding should inform the encoding. For instance, one may employ an ordinal encoding such as $\mathbf{u}_{1} = 1$, $\mathbf{u}_{2} = 3$, and $\mathbf{u}_{3} = 6$. Alternatively, a semicircle encoding like $\mathbf{u}_{1} = (1, 0)$, $\mathbf{u}_{2} = (1/2, \sqrt{3}/2)$, and $\mathbf{u}_{3} = (-1, 0)$ could be used so that $\| \mathbf{u}_{1} - \mathbf{u}_{2} \| < \| \mathbf{u}_{2} - \mathbf{u}_{3} \|$.

\subsection{Categorical distance correlation under general encodings}\label{sec22}
We now give the explicit expression of categorical distance correlation under general encodings. Let $X$ and $Y$ be two random variables, with categories $\{1,~ ...,~ I\}$ and $\{1,~ ...,~ J\}$, respectively. Given their encodings $\{\mathbf{u}_{1}, ..., \mathbf{u}_{I}\}$ and $\{\mathbf{v}_{1}, ..., \mathbf{v}_{J}\}$, we have the distance matrices $D^{X} = (d^{X}_{ik})_{1\leq i, k \leq I}$ and $D^{Y} = (d^{Y}_{jl})_{1\leq j, l \leq J}$, where $d^{X}_{ik}=\| \mathbf{u}_{i} - \mathbf{u}_{k} \|$ represents the Euclidean distance between categories $i$ and $k$, and $d^{Y}_{jl}=\| \mathbf{v}_{j} - \mathbf{v}_{l} \|$ between $j$ and $l$. Throughout this paper, without causing any confusion, we assume fixed encodings, and use the category $i$ and its unique encoding $\mathbf{u}_{i}$ exchangeably. 

Let $\pi = \{\pi_{ij}\}_{1\leq i\leq I, 1\leq j\leq J}$ be the joint distribution of $(X, Y)$, and $\{\pi_{i+}\}_{1\leq i\leq I}$ and $\{\pi_{+j}\}_{1\leq j\leq J}$ be the marginal probabilities. Let $n_{ij}$ be the observed count of $(X=i,~ Y=j)$, $n = \sum_{i=1}^{I}\sum_{j=1}^{J}n_{ij}$ be the total sample size, $n_{i+} = \sum_{j=1}^{J}n_{ij}$ and $n_{+j} = \sum_{i=1}^{I}n_{ij}$ be the marginal counts. The maximum likelihood estimates are $\widehat{\pi}_{ij} = n_{ij}/n$, $\widehat{\pi}_{i+} = n_{i+}/n$, and $\widehat{\pi}_{+j} = n_{+j}/n$. We use $\xrightarrow[]{d}$, $\xrightarrow[]{P}$, and $\xrightarrow[]{a.s.}$ to denote convergence in distribution, convergence in probability, and almost sure convergence, respectively. 

Lemmas \ref{dcov} and \ref{dvar} below give the expressions of the categorical distance covariance and variance, which extend\linkedyearcite{}{Zhang2019}{Zhang (2019)} and\linkedyearcite{}{USP}{Berrett \& Samworth (2021)} from one-hot encoding to general encodings (proofs are provided in \ref{A1} and \ref{A2}).
\begin{Lem}\label{dcov}
Under general encodings and distance matrices $D^{X}$ and $D^{Y}$, the squared distance covariance between categorical variables $X$ and $Y$ can be expressed as 
\begin{equation*}
\mathrm{dCov}^{2}(X, Y) = \sum_{i=1}^{I}\sum_{j=1}^{J}\sum_{k=1}^{I}\sum_{l=1}^{J}(\pi_{ij}-\pi_{i+}\pi_{+j})(\pi_{kl}-\pi_{k+}\pi_{+l})d^{X}_{ik}d^{Y}_{jl}.
\end{equation*}
An equivalent matrix form is 
\begin{equation*}
\mathrm{dCov}^{2}(X, Y) = \Delta^{T}(D^{X}\otimes D^{Y})\Delta,
\end{equation*}
where $D^{X}\otimes D^{Y}$ represents the Kronecker product of $D^{X}$ and $D^{Y}$, and $\Delta$ is the column vector $(\pi_{11}-\pi_{1+}\pi_{+1}, ..., \pi_{1J}-\pi_{1+}\pi_{+J}, ~...,~\pi_{IJ}-\pi_{I+}\pi_{+J})^{T}$. Specifically, for one-hot encodings, where $d^{X}_{ik} = \sqrt{2}I\{i\neq k\}$ and $d^{Y}_{jl} = \sqrt{2}I\{j\neq l\}$, we have $dCov^{2}(X, Y) = 2\sum_{i=1}^{I}\sum_{j=1}^{J}(\pi_{ij}-\pi_{i+}\pi_{+j})^{2}$.
\end{Lem}

\begin{Rem}
The Kronecker product is generally asymmetric, meaning $D^{X}\otimes D^{Y} \neq D^{Y}\otimes D^{X}$. In the lemma above, $\Delta$ is formed by row-wise vectorization. However, if $\Delta$ is defined using column-wise vectorization, i.e., $(\pi_{11}-\pi_{1+}\pi_{+1}, ..., \pi_{I1}-\pi_{I+}\pi_{+1}, ~...,~\pi_{IJ}-\pi_{I+}\pi_{+J})^{T}$, then $dCov^{2}(X, Y) = \Delta^{T}(D^{Y}\otimes D^{X})\Delta$.
\end{Rem}

\begin{Lem}\label{dvar}
Under general encodings and distance matrix $D^{X}$, the squared distance variance for categorical variable $X$ is
\begin{equation*}
\mathrm{dVar}^{2}(X) = \sum_{i=1}^{I}\sum_{k=1}^{I}\pi_{i+}\pi_{k+}(d^{X}_{ik}-\overline{d^{X}_{i}})(d^{X}_{ik}-\overline{d^{X}_{k}}),
\end{equation*}
where $\overline{d^{X}_{i}} = \sum_{k=1}^{I}\pi_{k+}d^{X}_{ik}$ represents the average distance between category $i$ and the others. Specifically, for one-hot encodings, we have $dVar^{2}(X) = 2\left( \sum_{i=1}^{I}\pi^{2}_{i+} \right)^{2} - 4\sum_{i=1}^{I}\pi^{3}_{i+} + 2\sum_{i=1}^{I}\pi^{2}_{i+}$.
\end{Lem}

Combining Lemmas \ref{dcov} and \ref{dvar}, we have the squared distance correlation under general encodings 
\begin{equation}\label{dcor}
\mathrm{dCor}^{2}(X, Y) = \frac{\sum_{i=1}^{I}\sum_{j=1}^{J}\sum_{k=1}^{I}\sum_{l=1}^{J}(\pi_{ij}-\pi_{i+}\pi_{+j})(\pi_{kl}-\pi_{k+}\pi_{+l})d^{X}_{ik}d^{Y}_{jl}}{\sqrt{\sum_{i=1}^{I}\sum_{k=1}^{I}\pi_{i+}\pi_{k+}(d^{X}_{ik}-\overline{d^{X}_{i}})(d^{X}_{ik}-\overline{d^{X}_{k}})}\sqrt{\sum_{j=1}^{J}\sum_{l=1}^{J}\pi_{+j}\pi_{+l}(d^{Y}_{jl}-\overline{d^{Y}_{j}})(d^{Y}_{jl}-\overline{d^{Y}_{l}})}}.
\end{equation} 

The above categorical distance correlation is scale-invariant with respect to $D^{X}$ and $D^{Y}$. This means that for any $c>0$, using $cD^{X}$ and $cD^{Y}$ yields the same $dCor^{2}(X, Y)$. For the remainder of this paper, we will use the rescaled distances such that $\max_{i,k}d^{X}_{ik} = \max_{j,l}d^{Y}_{jl} = 1$. For instance, with one-hot encodings, the rescaled distance matrix $D^{X}$ will contain zeros on the diagonal and ones elsewhere. This ensures that $dCov^{2}(X, Y)$, $dVar^{2}(X)$, and $dVar^{2}(Y)$ are all bounded, which simplifies our technical proofs.

We consider two sample estimates for the squared distance covariance, including the maximum likelihood estimate (MLE) and an unbiased estimate. The MLE can be obtained by replacing the true probabilities $\pi_{ij}$'s with their sample estimates $\widehat{\pi}_{ij}$'s
\begin{equation} \label{mle}
\widehat{\mathrm{dCov}^{2}}(X, Y) =  \sum_{i=1}^{I}\sum_{j=1}^{J}\sum_{k=1}^{I}\sum_{l=1}^{J}(\widehat{\pi}_{ij}-\widehat{\pi}_{i+}\widehat{\pi}_{+j})(\widehat{\pi}_{kl}-\widehat{\pi}_{k+}\widehat{\pi}_{+l})d^{X}_{ik}d^{Y}_{jl}.
\end{equation}

Motivated by\linkedyearcite{}{USP}{Berrett \& Samworth (2021)}, an unbiased estimate can be obtained using a fourth-order $U$-statistic. Let $(X_{1}, Y_{1}), ..., (X_{n}, Y_{n})$ be $i.i.d.$ samples, and $A_{gh} = \| X_{g} - X_{h}\|$, $B_{gh} = \| Y_{g} - Y_{h}\|$, where $\| X_{g} - X_{h}\|$ represents the Euclidean distance between $X_{g}$ and $X_{h}$. The unbiased estimate of $dCov^{2}$ is given by
\begin{equation} \label{mle}
\widetilde{\mathrm{dCov}^{2}}(X, Y) = \frac{T_{1}}{n(n-3)} -\frac{2T_{2}}{n(n-2)(n-3)}+\frac{T_{3}}{n(n-1)(n-2)(n-3)} , 
\end{equation}
where
\begin{align*}
T_{1} & = \sum_{g=1}^{n}\sum_{h=1}^{n}A_{gh}B_{gh} = \sum_{i=1}^{I}\sum_{j=1}^{J}\sum_{k=1}^{I}\sum_{l=1}^{J} n_{ij}n_{kl}d^{X}_{ik}d^{Y}_{jl} , \\
T_{2} & = \sum_{g=1}^{n}\left( \sum_{h=1}^{n}A_{gh}\sum_{h=1}^{n}B_{gh} \right) = \sum_{i=1}^{I}\sum_{j=1}^{J}\sum_{k=1}^{I}\sum_{l=1}^{J} n_{ij}n_{k+}n_{+l}d^{X}_{ik}d^{Y}_{jl}, \\
T_{3} & = \left( \sum_{g=1}^{n}\sum_{h=1}^{n}A_{gh} \right)\left( \sum_{g=1}^{n}\sum_{h=1}^{n}B_{gh} \right) = \sum_{i=1}^{I}\sum_{j=1}^{J}\sum_{k=1}^{I}\sum_{l=1}^{J} n_{i+}n_{+j}n_{k+}n_{+l}d^{X}_{ik}d^{Y}_{jl}. \\
\end{align*}

\begin{Rem}\label{mleV}
It can be shown that the MLE is essentially a $V$-statistic. Precisely, we have
\begin{equation*}
\widehat{\mathrm{dCov}^{2}}(X, Y) = \frac{T_{1}}{n^{2}} -\frac{2T_{2}}{n^{3}}+\frac{T_{3}}{n^{4}}.
\end{equation*}
\end{Rem}

\begin{Rem}\label{a.s.}
By the Strong Law of Large Numbers (SLLN), we have $\widehat{\pi}_{ij} \xrightarrow[]{a.s.} \pi_{ij}$. Further, by the continuous mapping theorem and the boundedness of the sample estimates under rescaled distances, we have 
\begin{align*}
\widehat{\mathrm{dCov}^{2}}(X, Y) & \xrightarrow[]{a.s.} \mathrm{dCov}^{2}(X, Y), \\
\widetilde{\mathrm{dCov}^{2}}(X, Y) & \xrightarrow[]{a.s.} \mathrm{dCov}^{2}(X, Y), \\
\widehat{\mathrm{dVar}^{2}}(X) & \xrightarrow[]{a.s.} \mathrm{dVar}^{2}(X), \\
\widetilde{\mathrm{dVar}^{2}}(X) & \xrightarrow[]{a.s.} \mathrm{dVar}^{2}(X).
\end{align*}
\end{Rem}

Lemma \ref{diff} and Remark \ref{inddiff} below give the asymptotic bias (scaled by $n$) of the MLE (proof is provided in \ref{A3}).
\begin{Lem} \label{diff}
Under general encodings and distance matrices $D^{X}$ and $D^{Y}$, we have 
\begin{equation*}
n \left[ \widehat{\mathrm{dCov}^{2}}(X, Y) - \widetilde{\mathrm{dCov}^{2}}(X, Y) \right]  \xrightarrow[]{a.s.}  \sum_{i=1}^{I}\sum_{j=1}^{J}\sum_{k=1}^{I}\sum_{l=1}^{J} (10\pi_{ij}\pi_{k+}\pi_{+l}-3\pi_{ij}\pi_{kl}-6\pi_{i+}\pi_{+j}\pi_{k+}\pi_{+l}) d^{X}_{ik}d^{Y}_{jl} . 
\end{equation*}
\end{Lem}

\begin{Rem} \label{inddiff}
Under the null hypothesis of independence, we have $\pi_{ij} = \pi_{i+}\pi_{+j}$, therefore the asymptotic bias of the MLE (scaled by $n$) becomes 
\begin{align*}
n\left[ \widehat{\mathrm{dCov}^{2}}(X, Y) - \widetilde{\mathrm{dCov}^{2}}(X, Y) \right] & \xrightarrow[]{a.s.} \sum_{i=1}^{I}\sum_{j=1}^{J}\sum_{k=1}^{I}\sum_{l=1}^{J} \pi_{i+}\pi_{+j}\pi_{k+}\pi_{+l}d^{X}_{ik}d^{Y}_{jl} \\
& = E(\|X_{1} - X_{2}\|)E(\|Y_{1} - Y_{2}\|),
\end{align*}
where $(X_{1}, Y_{1})$ and $(X_{2}, Y_{2})$ are two independent copies of $(X, Y)$, and $E(\|X_{1} - X_{2}\|)$ is the average distance between two categories of $X$. 
\end{Rem}

Similarly, for the squared distance variance, we have the following estimates
\begin{align*} 
\widetilde{\mathrm{dVar}^{2}}(X, Y) & = \frac{T_{1}}{n(n-3)} -\frac{2T_{2}}{n(n-2)(n-3)}+\frac{T_{3}}{n(n-1)(n-2)(n-3)} , \\
\widehat{\mathrm{dVar}^{2}}(X, Y) & = \frac{T_{1}}{n^2} -\frac{2T_{2}}{n^3}+\frac{T_{3}}{n^4} ,
\end{align*}
where
\begin{align*}
T_{1} & = \sum_{g=1}^{n}\sum_{h=1}^{n}A_{gh}^{2} = \sum_{i=1}^{I}\sum_{k=1}^{I} n_{i+}n_{k+}\left( d^{X}_{ik} \right)^{2}, \\
T_{2} & = \sum_{g=1}^{n}\left( \sum_{h=1}^{n}A_{gh}\right)^{2} = \sum_{i=1}^{I}\sum_{k=1}^{I}\sum_{m=1}^{I} n_{i+}n_{k+}n_{m+}d^{X}_{ik}d^{X}_{im}, \\
T_{3} & = \left( \sum_{g=1}^{n}\sum_{h=1}^{n}A_{gh} \right)^{2} = \left( \sum_{i=1}^{I}\sum_{k=1}^{I} n_{i+}n_{k+}d^{X}_{ik} \right)^{2}. 
\end{align*}
Similar to Lemma \ref{diff}, it is straightforward to show  
\begin{equation*}
n\left[\widehat{\mathrm{dVar}^{2}}(X) - \widetilde{\mathrm{dVar}^{2}}(X)\right] \xrightarrow[]{a.s.} \sum_{i=1}^{I}\sum_{k=1}^{I} \pi_{i+}\pi_{k+}\left[ 10d^{X}_{ik}\overline{d^{X}_{i}}-3\left( d^{X}_{ik}\right)^{2}-6\overline{d^{X}_{i}}~\overline{d^{X}_{k}} \right].
\end{equation*}

\subsection{Asymptotic distributions}\label{sec23}
In this part, we give the asymptotic distributions of the two estimates in \ref{sec22}.  We first derive their asymptotic null distributions. To begin, let $\mathbf{1}_{n}$ be the $n$-by-$n$ identity matrix, and $\mathbf{1}_{n\times n}$ be the $n$-by-$n$ matrix with all elements equal to 1. Define the centering matrix $H_{n} = \mathbf{1}_{n} - \frac{1}{n}\mathbf{1}_{n\times n}$. For the bias-corrected estimate of squared distance correlation, by\linkedyearcite{}{Zhangetal}{Zhang et al. (2018)}, we have 
\begin{equation*}
n\widetilde{\mathrm{dCor}^{2}}(X, Y) \xrightarrow{d} \frac{1}{\sqrt{\sum_{g=1}^{\infty}\lambda^{2}_{g}\sum_{h=1}^{\infty}\mu^{2}_{h}}}\sum_{i=1}^{\infty}\sum_{j=1}^{\infty} \lambda_{i}\mu_{j}(Z^{2}_{ij}-1),
\end{equation*}
as $n\rightarrow\infty$, where $Z_{ij}$'s are independent standard normal random variables. The weights $\{\lambda_{i}\}$ and $\{\mu_{j}\}$ are the limiting eigenvalues of $H_{n}D_{n}^{X}H_{n}/n$ and $H_{n}D_{n}^{Y}H_{n}/n$, respectively, where $D_{n}^{X}$ and $D_{n}^{Y}$ are the sample distance matrices, i.e., the element $(g, h)$ in $D_{n}^{X}$ is $\|X_{g} - X_{h}\|$.

For continuous data, calculating $\{\lambda_{i}\}$ and $\{\mu_{j}\}$ generally requires solving the eigenvalues of two $n\times n$ matrices, which can be computationally expensive when $n$ is large. This cost, however, is significantly reduced for categorical variables, making the approximation feasible. The theorem below establishes the asymptotic null distribution of the bias-corrected estimate, requiring only the calculation of eigenvalues from an $I\times I$ and a $J\times J$ matrix (proof in \ref{A4}).

\begin{Thm}\label{null}
Under the null hypothesis of independence, and general encodings with distance matrices $D^{X}$ and $D^{Y}$, we have 
\begin{equation*}
n\widetilde{\mathrm{dCor}^{2}}(X, Y) \xrightarrow{d} \frac{1}{\sqrt{\sum_{g=1}^{I-1}\lambda^{2}_{g}\sum_{h=1}^{J-1}\mu^{2}_{h}}}\sum_{i=1}^{I-1}\sum_{j=1}^{J-1} \lambda_{i}\mu_{j}(Z^{2}_{ij}-1),
\end{equation*}
where $\{\lambda_{1}, ..., \lambda_{I-1}\}$ and $\{\mu_{1}, ..., \mu_{J-1}\}$ are the eigenvalues of the following matrices
  \begin{equation*}
  Q^{X} = 
 \begin{bmatrix} 
    -\overline{d^{X}_{1}}\pi_{1+} & (d^{X}_{12}-\overline{d^{X}_{1}})\sqrt{\pi_{1+}\pi_{2+}} & \dots & (d^{X}_{1I}-\overline{d^{X}_{1}})\sqrt{\pi_{1+}\pi_{I+}} \\
   (d^{X}_{21}-\overline{d^{X}_{2}})\sqrt{\pi_{2+}\pi_{1+}} &  -\overline{d^{X}_{2}}\pi_{2+} & \dots & (d^{X}_{2I}-\overline{d^{X}_{2}})\sqrt{\pi_{2+}\pi_{I+}} \\
    \vdots & \vdots  & \ddots & \vdots \\
    (d^{X}_{I1}-\overline{d^{X}_{I}})\sqrt{\pi_{I+}\pi_{1+}} &  (d^{X}_{I2}-\overline{d^{X}_{I}})\sqrt{\pi_{I+}\pi_{2+}} & \dots &  -\overline{d^{X}_{I}}\pi_{I+} 
    \end{bmatrix},
 \end{equation*}  
   \begin{equation*}
    Q^{Y} = 
 \begin{bmatrix} 
    -\overline{d^{Y}_{1}}\pi_{+1} & (d^{Y}_{12}-\overline{d^{Y}_{1}})\sqrt{\pi_{+1}\pi_{+2}} & \dots & (d^{Y}_{1J}-\overline{d^{Y}_{1}})\sqrt{\pi_{+1}\pi_{+J}} \\
   (d^{Y}_{21}-\overline{d^{Y}_{2}})\sqrt{\pi_{+2}\pi_{+1}} &  -\overline{d^{Y}_{2}}\pi_{+2} & \dots & (d^{Y}_{2J}-\overline{d^{Y}_{2}})\sqrt{\pi_{+2}\pi_{+J}} \\
    \vdots & \vdots  & \ddots & \vdots \\
    (d^{Y}_{J1}-\overline{d^{Y}_{J}})\sqrt{\pi_{+J}\pi_{+1}} &  (d^{Y}_{J2}-\overline{d^{Y}_{J}})\sqrt{\pi_{+J}\pi_{+2}} & \dots &  -\overline{d^{Y}_{J}}\pi_{+J} 
    \end{bmatrix}.
 \end{equation*}  
\end{Thm}

Under one-hot encodings for nominal $X$ and $Y$, the two matrices in Theorem \ref{null} simplify to 
\begin{equation*}
  Q^{X} = 
 \begin{bmatrix} 
    (\pi_{1+}-1)\pi_{1+} & \pi_{1+}\sqrt{\pi_{1+}\pi_{2+}} & \dots & \pi_{1+}\sqrt{\pi_{1+}\pi_{I+}} \\
   \pi_{2+}\sqrt{\pi_{2+}\pi_{1+}} &  (\pi_{2+}-1)\pi_{2+} & \dots & \pi_{2+}\sqrt{\pi_{2+}\pi_{I+}} \\
    \vdots & \vdots  & \ddots & \vdots \\
    \pi_{I+}\sqrt{\pi_{I+}\pi_{1+}} &  \pi_{I+}\sqrt{\pi_{I+}\pi_{2+}} & \dots &  (\pi_{I+}-1)\pi_{I+}
    \end{bmatrix},
 \end{equation*}  
   \begin{equation*}
    Q^{Y} = 
\begin{bmatrix} 
    (\pi_{+1}-1)\pi_{+1} & \pi_{+1}\sqrt{\pi_{+1}\pi_{+2}} & \dots & \pi_{+1}\sqrt{\pi_{+1}\pi_{+J}} \\
   \pi_{+2}\sqrt{\pi_{+2}\pi_{+1}} &  (\pi_{+2}-1)\pi_{+2} & \dots & \pi_{+2}\sqrt{\pi_{+2}\pi_{+J}} \\
    \vdots & \vdots  & \ddots & \vdots \\
    \pi_{+J}\sqrt{\pi_{+J}\pi_{+1}} &  \pi_{+J}\sqrt{\pi_{+J}\pi_{+2}} & \dots &  (\pi_{+J}-1)\pi_{+J}
    \end{bmatrix}.
 \end{equation*}  

By Theorem \ref{null}, Lemma \ref{diff} and Slutsky's theorem, we can obtain the asymptotic null distribution of the MLE, as shown in the following corollary
\begin{Cor} \label{bias.corrected}
Under the null hypothesis of independence, and general encodings with distance matrices $D^{X}$ and $D^{Y}$, we have
\begin{equation*}
n\widehat{\mathrm{dCor}^{2}}(X, Y) \xrightarrow{d} \frac{1}{\sqrt{\sum_{g=1}^{I-1}\lambda^{2}_{g}\sum_{h=1}^{J-1}\mu^{2}_{h}}} \left [\sum_{i=1}^{I-1}\sum_{j=1}^{J-1} \lambda_{i}\mu_{j}(Z^{2}_{ij}-1) + B \right ],
\end{equation*}
where
\begin{equation*}
B = \sum_{i=1}^{I}\sum_{j=1}^{J}\sum_{k=1}^{I}\sum_{l=1}^{J} \pi_{i+}\pi_{+j}\pi_{k+}\pi_{+l} d^{X}_{ik}d^{Y}_{jl},
\end{equation*}
and $\{\lambda_{1}, ..., \lambda_{I-1}\}$ and $\{\mu_{1}, ..., \mu_{J-1}\}$ are defined same as in Theorem \ref{null}.
\end{Cor}

Applying Theorem \ref{null} and Corollary \ref{bias.corrected} for p-value calculations requires estimating the marginal probabilities and the average distances in matrices $Q^{X}$ and $Q^{Y}$, followed by solving the eigenvalues $\{\lambda_{1}, ..., \lambda_{I-1}\}$ and $\{\mu_{1}, ..., \mu_{J-1}\}$. While the weighted sum of chi-squared random variables in general does not have a closed-form $c.d.f.$, numerous statistical packages, such as R function \texttt{psum.chisq} from package \texttt{mgcv} by\linkedyearcite{}{Wood}{Wood (2025)}, provide accurate and efficient approximations.

Next, we give the asymptotic distributions of the two estimates under fixed alternatives. Let $\mathrm{vec}(A)$ denote the vectorization of matrix $A$, obtained by stacking all the columns. Let $\Sigma$ be the variance-covariance matrix of $\mathrm{vec}(\sqrt{n}\widehat{\pi})$, where $\mathrm{Var}(\sqrt{n}\widehat{\pi}_{ij}) = \pi_{ij}(1-\pi_{ij})$ and $\mathrm{Cov}(\sqrt{n}\widehat{\pi}_{ij},~\sqrt{n}\widehat{\pi}_{km}) =  - \pi_{ij} \pi_{km}$ for $i\neq k$ or $j\neq m$. For instance, for $I=3$ and $J=4$, $\Sigma$ is a 12-by-12 matrix. In addition, let $D' = (D'_{ij})$ be the first-order partial derivatives of $dCov^{2}(X, Y)$, i.e., $D'_{ij} = \frac{\partial}{\partial \pi_{ij}}dCov^{2}(X, Y)$. The following theorem establishes the asymptotic normality of the two estimates under fixed alternatives. The proof is based on multivariate delta method, as detailed in \ref{A5}.

\begin{Thm}\label{alter}
Under fixed alternatives, and general encodings with distance matrices $D^{X}$ and $D^{Y}$, we have
\begin{align*}
\sqrt{n}(\widehat{\mathrm{dCor}^{2}}(X, Y) - \mathrm{dCor}^{2}(X, Y)) & \xrightarrow{d} N\left\{ 0, ~ \frac{[ \mathrm{vec}(D') ]^{\intercal} \Sigma \mathrm{vec}(D')}{\sum_{i=1}^{I-1}\lambda^{2}_{i}\sum_{j=1}^{J-1}\mu^{2}_{j}} \right\}, \\
\sqrt{n}(\widetilde{\mathrm{dCor}^{2}}(X, Y) - \mathrm{dCor}^{2}(X, Y)) & \xrightarrow{d} N\left\{ 0, ~ \frac{[ \mathrm{vec}(D') ]^{\intercal} \Sigma \mathrm{vec}(D')}{\sum_{i=1}^{I-1}\lambda^{2}_{i}\sum_{j=1}^{J-1}\mu^{2}_{j}} \right\}, 
\end{align*}
where  
\begin{equation*}
D'_{ij} = \sum_{k=1}^{I}\sum_{l=1}^{J} (\pi_{kl} - \pi_{k+}\pi_{+l}) \left(d^{X}_{ik}d^{Y}_{jl} - d^{X}_{ik}\overline{d^{Y}_{l}} - d^{Y}_{jl}\overline{d^{X}_{k}} \right),
\end{equation*}
and $\{\lambda_{1}, ..., \lambda_{I-1}\}$ and $\{\mu_{1}, ..., \mu_{J-1}\}$ are defined same as in Theorem \ref{null}.
\end{Thm}

Under independence, all $D'_{ij}$'s are zero, which results in an asymptotic variance of zero in Theorem \ref{alter}. Consequently, $\sqrt{n}\widehat{dCor^{2}}(X, Y)$ and $\sqrt{n}\widetilde{dCor^{2}}(X, Y)$ are both degenerate, converging to a weight sum of chi-squared distributions (see Theorem \ref{null} and Corollary \ref{bias.corrected}). 

\subsection{Application to high-dimensional feature screening}
In this section, we apply the categorical distance correlation to high-dimensional feature screening\linkedyearcite{}{FanLv}{(Fan \& Lv, 2008};\linkedyearcite{}{FanSong}{Fan \& Song, 2010};\linkedyearcite{}{LiZhongZhu}{Li et al., 2012)}. For categorical data,\linkedyearcite{}{Huangetal}{Huang et al. (2014)} proposed a sure screening procedure based on Pearson's chi-squared divergence. More recently,\linkedyearcite{}{Zhang2025}{Zhang (2025)} investigated categorical distance covariance for feature screening, demonstrating its consistency under slightly weaker conditions than\linkedyearcite{}{Huangetal}{Huang et al. (2014)}. A common limitation of\linkedyearcite{}{Huangetal}{Huang et al. (2014)} and\linkedyearcite{}{Zhang2025}{Zhang (2025)}, however, is that their methods are only suitable for nominal variables, potentially leading to information loss when dealing with ordinal data. To address this, we apply the categorical distance correlation derived in \ref{sec22} to handle any categorical variables with general encodings.

We begin with the notations and problem formulation. Let $Y\in\{1,~ ...,~ J\}$ be the response variable with distance matrix $D^{Y}$, and $\mathcal{S} = \{X^{1},~ ...,~ X^{S}\}$ be the set of $S$ categorical candidate features of any type, e.g., $\mathcal{S}$ can be a mix of nominal and ordinal variables. Suppose feature $X^{s}$ has sample space $\{1,~ ...,~ I_{s}\}$ and distance matrix $D^{X^{s}}$ (determined by its encodings), we let $\pi_{i_{s}j} = P(X^{s} = i_{s}, Y = j)$ and $\pi_{i_{s}+} = P(X^{s} = i_{s})$ denote the joint and marginal probabilities. Let $(X^{1}_{m},~ ...,~ X^{S}_{m})$, $m = 1,~ ...,~ n$, be $n$ independent samples, then $\pi_{i_{s}j}$ and $\pi_{i_{s}+}$ can be estimated by corresponding sample proportions, denoted by $\widehat{\pi}_{i_{s}j}$ and $\widehat{\pi}_{i_{s}+}$. Finally, we define the true model, denoted by $\mathcal{S}_{T}$, as a non-empty subset of features in $\mathcal{S}$, such that $X^{s}\in \mathcal{S}_{T}$ if and only if $X^{s}$ and $Y$ are dependent.

To measure the dependence between $X^{s}$ and $Y$, we consider the following maximum likelihood estimate of $dCor^{2}(X^{s}, Y)$ (the bias-corrected estimate can be established in the same way) 
\begin{equation*} 
\widehat{\mathrm{dCor}^{2}}(X^{s}, Y) =  \sum_{i_{s}=1}^{I_{s}}\sum_{j=1}^{J}\sum_{k_{s}=1}^{I_{s}}\sum_{l=1}^{J}(\widehat{\pi}_{i_{s}j}-\widehat{\pi}_{i_{s}+}\widehat{\pi}_{+j})(\widehat{\pi}_{k_{s}l}-\widehat{\pi}_{k_{s}+}\widehat{\pi}_{+l})d^{X}_{i_{s}k_{s}}d^{Y}_{jl}.
\end{equation*}
Obviously, features with larger value of $\widehat{dCor}^{2}(X^{s}, Y)$ are more likely to be relevant, therefore we can estimate the true model $\mathcal{S}_{T}$ by $\widehat{S}(C) = \{X^{s}:~ \widehat{dCor}^{2}(X^{s}, Y)>C \}$, where $C>0$ is some predefined constant. We demonstrate that, under certain conditions, this method can consistently identify the true model $\mathcal{S}_{T}$. Our screening relies on the following conditions
\begin{itemize}
\item[(C1)] There exists a positive constant $I_{max}$, such that $\max_{X^{s}\in \mathcal{S}}I_{s} < I_{max}$.
\item[(C2)] There exists positive constants, $\delta^{2}_{min}>0$ and $\sigma^{2}_{min}>0$, such that $\min_{X^{s}\in \mathcal{S}_{T}} dCov^{2}(X^{s}, Y)>\delta^{2}_{min}$ and $dVar(Y)\wedge\min_{X^{s}\in \mathcal{S}} dVar(X^{s}) > \sigma^{2}_{min}$. 
\item[(C3)] $\log S = o(n)$.
\end{itemize}
Condition (C1) assumes that the number of categories for all features is finite, as dealing with a diverging number of categories in feature screening is theoretically challenging. For example,\linkedyearcite{}{Huangetal}{Huang et al. (2014)} and\linkedyearcite{}{Zhang2025}{Zhang (2025)} also assumed bounded $\max_{X^{s}\in \mathcal{S}}I_{s}$. Condition (C2) guarantees lower bounds for the distance covariance and distance variance of all relevant features (those in $\mathcal{S}_{T}$), which allow them to be asymptotically distinguishable from irrelevant features. Condition (C3) requires that the feature dimension $S$ diverges no faster than the exponential of sample size. 

Lemma \ref{concentrate} below establishes the uniform consistency of $\widehat{dCor^{2}}(X^{s}, Y)$ (proof in \ref{A6}) 
\begin{Lem}\label{concentrate}
Under Conditions (C1)-(C3), for any $\epsilon > 0$, we have
\begin{equation*}
P\left(\max_{X^{s}\in \mathcal{S}} \left | \widehat{\mathrm{dCor}^{2}}(X^{s}, Y) - \mathrm{dCor}^{2}(X^{s}, Y) \right | >\epsilon \right) \leq  2I_{max} J\cdot\exp\left[ \log S-\frac{6n\epsilon^2}{4\epsilon I_{max}J \left(  \frac{4}{\sigma^{8}_{min}} + \frac{2}{\sigma^{10}_{min}} \right) + 3I^{2}_{max}J^{2} \left(  \frac{4}{\sigma^{8}_{min}} + \frac{2}{\sigma^{10}_{min}} \right)^{2}} \right].
\end{equation*}
\end{Lem}

By Lemma \ref{concentrate}, we can establish the sure screening property (proof of Theorem \ref{consistent} is given in \ref{A7})
\begin{Thm}\label{consistent}
Under Conditions (C1)-(C3), there exists a positive constant $C$, such that 
\begin{align*}
\lim_{n\rightarrow \infty}P\left[\widehat{\mathcal{S}}(C) = \mathcal{S}_{T}\right] = 1.
\end{align*}
\end{Thm}

To select the tuning parameter $C$, we employed a data-driven approach based on change-point detection introduced by\linkedyearcite{}{Zhang2025}{Zhang (2025)}. This method has demonstrated good performance in estimating the number of relevant features, particularly under relatively small sample sizes. The change-point based procedure begins by ordering all features in descending order according to their squared distance correlation with the response variable, i.e., $\widehat{dCor}^{2}(X^{(S)}, Y)\geq \cdots \geq \widehat{dCor}^{2}(X^{(1)}, Y)$. As the relevant features ($X^{s}\in\mathcal{S}_{T}$) are anticipated to have stronger sample distance correlation compared to those irrelevant ones ($X^{s}\in\mathcal{S}\setminus\mathcal{S}_{T}$), this distributional difference is expected to yield a change point where the slope of the ordered sequence undergoes a significant change. This change point serves as the threshold for discriminating between relevant and irrelevant features, thereby providing an estimate for $C$. We utilized a two-piece linear model, which can be conveniently implemented using the \texttt{segmented} package\linkedyearcite{}{Muggeo}{(Muggeo, 2024)}. Our simulation studies indicated that this approach is robust in selecting the appropriate tuning parameter $C$.

Our proposed feature screening procedure necessitates the pre-specification of encodings (or distance matrices) for all categorical variables, including the response and all candidate features. For nominal variables, one-hot encoding is standard given the absence of a natural ordering among their categories. However, for ordinal variables, there are several encoding options, including one-hot encoding (which disregards inherent order), ordinal encoding, and semicircle encoding, as detailed in \ref{sec21}. While a definitive answer for optimal encoding remains elusive, our simulation studies demonstrated that semicircle encoding yields overall satisfactory performance compared to one-hot and ordinal encodings. In practical applications, when variable-specific or manual encodings are infeasible, we recommend semicircle encoding for ordinal variables.

\section{Simulation study}\label{sec3}
In this section, we conduct a simulation study to evaluate the proposed feature screening procedure. In particular, we compare the performance of three encodings (one-hot, equally spaced ordinal, and equally spaced semicircle) for ordinal variables under high-dimensional settings, where the sample size $n$ is smaller than the number of candidate features $S$. In all settings, we use $S=10,000$, $|\mathcal{S}_{T}|=500$ (only 5\% of the features are relevant), and four sample sizes $n=\{25, 50, 75, 100\}$. By Equation \ref{dcor}, the magnitude of distance correlation is determined by the encodings and the strength of dependence, measured by $(\pi_{i_{s}j} - \pi_{i_{s}+}\pi_{+j})_{1\leq i_{s}\leq I_{s}, 1\leq j\leq J}$. The parameter setup for the relevant features ($X^{s}\in\mathcal{S}_{T}$) under each simulation setting is detailed below, and the dependence strength matrices $(\pi_{i_{s}j} - \pi_{i_{s}+}\pi_{+j})_{1\leq i_{s}\leq I_{s}, 1\leq j\leq J}$ are visualized in Figure 1. For irrelevant features ($X^{s}\in\mathcal{S}\setminus\mathcal{S}_{T}$), we use the same marginal distributions as those relevant features but $\pi_{i_{s}j} = \pi_{i_{s}+}\pi_{+j}$.

\begin{itemize}
\item Setting 1: $I_{s} = J = 5$ with marginal probabilities $\{0.5, 0.3, 0.1, 0.05, 0.05\}$ and a \underline{linear} dependence pattern, where $\pi_{i_{s}j} - \pi_{i_{s}+}\pi_{+j}=0.04$ for $(i_{s},j)\in\{(1,1), (2,2), (3,3), (4,4), (5,5)\}$. 
\item Setting 2: Same as Setting 1, but with a \underline{nonlinear and monotone} dependence pattern, where $\pi_{i_{s}j} - \pi_{i_{s}+}\pi_{+j}=0.04$ for $(i_{s},j)\in\{(1,1), (1,2), (2,3), (3,4), (4,4), (5,5)\}$.
\item Setting 3: Same as Setting 1, but with a \underline{nonmonotone} dependence pattern, where $\pi_{i_{s}j} - \pi_{i_{s}+}\pi_{+j}=0.04$ for $(i_{s},j)\in\{(1,1), (2,2), (3,3), (4,3), (2,4), (1,5)\}$.
\item Setting 4: $I_{s} = J = 8$ with marginal probabilities $\{0.5, 0.15, 0.1, 0.1, 0.05, 0.05, 0.03, 0.02\}$, and a \underline{linear} dependence pattern, where $\pi_{i_{s}j} - \pi_{i_{s}+}\pi_{+j}=0.025$ for $(i_{s},j)\in\{(1,1), (2,2), ... , (8,8)\}$.
\item Setting 5: Same as Setting 4, but with a \underline{nonlinear and monotone} dependence pattern, where $\pi_{i_{s}j} - \pi_{i_{s}+}\pi_{+j}=0.025$ for $(i_{s},j)\in\{(1,1), (1,2), (2,3), (3,4), (4,5), (4,6), (5,7), (6,7), (7,7), (8,8)\}$.
\item Setting 6: Same as Setting 4, but with a \underline{nonmonotone} dependence pattern, where $\pi_{i_{s}j} - \pi_{i_{s}+}\pi_{+j}=0.025$ for $(i_{s},j)\in\{(2,1), (3,1), (4,2), (5,3), (6,4), (6,5), (5,6), (4,7), (3,8), (2,8)\}$.
\end{itemize}

\begin{center}
[Figure 1 about here]
\end{center}

Figures 2-7 display the ROC curves for three encodings across various simulation settings and sample sizes. The AUCs are summarized in Tables 1 and 2, which also present the sensitivity and specificity based on the tuning parameters determined by the change-point approach. Despite the high dimensions and relatively small sample sizes, the distance correlation screenings generally perform well. For instance, in Setting 4 with $n=25$ (average cell count 0.39), the three screenings achieve AUCs of 0.742 (one-hot), 0.767 (ordinal), and 0.768 (semicircle). Furthermore, in the four monotone settings (1, 2, 4 and 5), the ordinal and semicircle encodings consistently achieve higher (3\% to 6\%) AUCs than the one-hot encoding (e.g., Setting 2 with $n=50$, one-hot 0.859, ordinal 0.903 and semicircle 0.906). However, in the two nonmonotone settings (3 and 6), the one-hot encoding achieves the highest AUCs, and the semicircle encoding consistently outperforms the ordinal encoding (e.g., Setting 6 with $n=100$, one-hot 0.934, ordinal 0.852, semicircle 0.894). In most settings, the change-point selected tuning parameter yields reasonable sensitivities and specificities, such as in Setting 5 with $n=75$ (average cell count 1.17), where the ordinal and semicircle encodings yield sensitivities of 0.743 and 0.746, and specificities of 0.982 and 0.980.

\begin{center}
[Figures 2-7 about here]
\end{center}

\begin{center}
[Table 1 about here]
\end{center}

\begin{center}
[Table 2 about here]
\end{center}

Our simulation study offers some insights into the performance of different encoding strategies for ordinal variables within the distance correlation screening framework. Particularly, the underlying relationship between ordinal variables impacts the optimal encoding choice. Generally, when two ordinal variables exhibit a linear or monotone trend, which is frequently observed in real-world datasets, the ordinal and semicircle encodings achieve better classification performance than the one-hot encoding. This is likely due to their ability to leverage the inherent order information that the one-hot encoding loses. Conversely, when ordinal variables exhibit a nonmonotone trend, the one-hot encoding demonstrates better sensitivity to detect these relationships. This suggests that in the presence of nonmonotone dependence patterns, treating categories as unordered, as in one-hot encoding, becomes more advantageous. While the semicircle encoding also yielded good classification results in these nonmonotone settings, the ordinal encoding is generally not recommended due to its lower AUC. Considering overall performance and practical applicability, especially when manual or variable-specific encoding is infeasible, we recommend the semicircle encoding for ordinal variables as a robust choice for our distance correlation screenings.

\section{A real application}\label{sec4}
We demonstrate the practical utility of the proposed distance correlation screening using the General Social Survey (GSS) data. The GSS, a nationally representative survey of U.S. adults, has served as a foundational resource for empirical sociological research for over five decades. This comprehensive dataset offers valuable insights into the evolving landscape of American society, documenting shifts in social attitudes, behaviors, and demographic patterns. For instance, in our prior investigation\linkedyearcite{}{YangZhangAngton}{(Yang et al., 2024)}, we employed GSS data to examine the association between a key measure of people's belief in meritocracy and other categorical features.

For this analysis, we focus on the 2018 GSS data, with a sample size of 2,348 respondents. The majority of the variables (802 out of 1,065) are categorical, which contains 166 nominal variables (e.g., race, gender) and 636 ordinal variables (e.g., education level, socioeconomic status). Our primary response variable is the subjective socioeconomic classification, which is ordinal with four groups: upper class, middle class, working class, and lower class. Our objective is to identify variables significantly associated with individuals' subjective socioeconomic classification, thereby offering insights into the disparities between these socioeconomic strata.

After data preprocessing, which involves excluding missing values and continuous variables, our final dataset includes 2,333 respondents. We then apply a distance correlation screening procedure to identify factors important to the response variable. Drawing upon findings from our simulation study, we employ one-hot encoding for all nominal variables and equally spaced semicircle encoding for all ordinal variables to optimize classification performance. Our change-point based method identifies a cutoff value of 0.031 (squared distance correlation), which yields a set of 32 relevant features out of the initial 802 candidate variables (four variables were excluded due to duplications). The identified list of relevant features, detailed in Table 3, provides substantial information regarding the distinctions among the socioeconomic classes. Notably, several variables related to education background, such as the highest degree attained and college-level courses completed, exhibit strong associations with the subjective socioeconomic classification. Furthermore, variables concerning the education levels of close family members, specifically spouse/partner's highest degree and father's highest degree, show significant differences across the four classes. Individuals' social networks are also connected to their subjective socioeconomic class, including whether the respondent knows an executive at a large company (KNWEXEC) or a lawyer (KNWLAWYR). Other important categorical features correlated with the subjective socioeconomic class include general health status and the perceived level of physical effort required in daily life.

\begin{center}
[Table 3 about here]
\end{center}

\section{Discussion and conclusions}\label{sec5}
The distance correlation by\linkedyearcite{}{DC}{Sz\'{e}kely et al. (2007)} is a powerful tool to detect and measure the association between two random vectors of arbitrary dimensions. As a special case, the categorical distance correlation offers many advantages over Pearson's chi-squared test \linkedyearcite{}{Zhang2025b}{(Zhang, 2025b};\linkedyearcite{}{Zhang2026}{Zhang, 2026};\linkedyearcite{}{ZhangDu2019}{Zhang \& Du, 2019)}. However, current methods are limited to one-hot encoding, therefore only suitable for nominal data. This paper generalizes the concept of categorical distance correlation to accommodate general categorical variables with arbitrary encodings. We establish some key properties including the asymptotic distributions and high-dimensional screening consistency. This significantly expands the applicability of the categorical distance correlation, making it readily usable with various real-world datasets, such as the well-known GSS data.   

There are several possible extensions of this work. First, while we have primarily utilized Euclidean distance to define spacings between categories ($D^{X}$ and $D^{Y}$) for illustrative purposes, it is noteworthy that distance correlation maintains its consistency (i.e., it is zero if and only if independence holds) across any metric space of strong negative type\linkedyearcite{}{Lyons13}{(Lyons, 2013};\linkedyearcite{}{Shen20}{Shen \& Vogelstein, 2020)}. It would be interesting to explore and evaluate the performance of certain non-Euclidean distances, such as the Gaussian kernel induced distances.

Second, our theoretical findings in this paper rely on the assumption of fixed encoding or predetermined between-category distances, meaning that the choice of encoding cannot be random. A possible extension would be investigating the properties of categorical distance correlation under data-dependent encodings ($D^{X}$ and $D^{Y}$ are functions of $\widehat{\pi}_{ij}$'s), such as frequency encoding and midranks\linkedyearcite{}{Agresti}{(Agresti, 2006)}. This could lead to a more adaptive and robust analytical framework.

Third, for the purpose of feature screening, it would be advantageous to incorporate more sophisticated methodologies for tuning parameter selection to control false discovery rate (FDR) or familywise error rate (FWER). For instance,\linkedyearcite{}{Guo}{Guo et al. (2023)} introduced a data-adaptive threshold selection procedure with error rate control, which is applicable to a broad range of popular screening methods, including our distance correlation screening. Their main idea involves employing a data-splitting strategy to construct a series of statistics possessing a marginal symmetry property, which is then leveraged to approximate the number of false discoveries.

Lastly, when applying our feature screening procedure to sensitive survey datasets, it is crucial to ensure respondent privacy. For this reason, an intriguing extension would be developing a differentially private version of our screening procedure, following the principles outlined by\linkedyearcite{}{Dwork}{Dwork (2006)} and\linkedyearcite{}{DP}{Wasserman \& Zhou (2010)}. This would allow us to protect individual privacy while preserving information content.

\section{Appendix}
\subsection{Proof of Lemma \ref{dcov}}\label{A1}
Let $(X_{1}, Y_{1})$, $(X_{2}, Y_{2})$ and $(X_{3}, Y_{3})$ be three independence copies of $(X, Y)$. By elementary probability, we have 
\begin{align*}
E(\|X_{1} - X_{2}\|) & = \sum_{i=1}^{I}\sum_{k=1}^{I}\pi_{i+}\pi_{k+}d^{X}_{ik} , \\
E(\|Y_{1} - Y_{2}\|) & = \sum_{j=1}^{J}\sum_{l=1}^{J}\pi_{+j}\pi_{+l}d^{Y}_{jl} , \\
E(\|X_{1} - X_{2}\|\|Y_{1} - Y_{2}\|) & =  \sum_{i=1}^{I}\sum_{j=1}^{J}\sum_{k=1}^{I}\sum_{l=1}^{J} \pi_{ij}\pi_{kl}d^{X}_{ik}d^{Y}_{jl}, \\
E(\|X_{1} - X_{2}\|\|Y_{1} - Y_{3}\|) & =  \sum_{i=1}^{I}\sum_{j=1}^{J}\sum_{k=1}^{I}\sum_{l=1}^{J} \pi_{ij}\pi_{k+}\pi_{+l}d^{X}_{ik}d^{Y}_{jl}.
\end{align*}
By the definition of $dCov^{2}(X, Y)$, we have 
\begin{align*}
\mathrm{dCov^{2}(X, Y)} = & E(\|X_{1} - X_{2}\|\|Y_{1} - Y_{2}\|) + E(\|X_{1} - X_{2}\|)E(\|Y_{1} - Y_{2}\|) - 2E(\|X_{1} - X_{2}\|\|Y_{1} - Y_{3}\|) \\
= &  \sum_{i=1}^{I}\sum_{j=1}^{J}\sum_{k=1}^{I}\sum_{l=1}^{J} \pi_{ij}\pi_{kl}d^{X}_{ik}d^{Y}_{jl} +  \sum_{i=1}^{I}\sum_{j=1}^{J}\sum_{k=1}^{I}\sum_{l=1}^{J} \pi_{i+}\pi_{+j}\pi_{k+}\pi_{+l}d^{X}_{ik}d^{Y}_{jl} \\
& - \sum_{i=1}^{I}\sum_{j=1}^{J}\sum_{k=1}^{I}\sum_{l=1}^{J} \pi_{ij}\pi_{k+}\pi_{+l}d^{X}_{ik}d^{Y}_{jl} - \sum_{i=1}^{I}\sum_{j=1}^{J}\sum_{k=1}^{I}\sum_{l=1}^{J} \pi_{kl}\pi_{i+}\pi_{+j}d^{X}_{ik}d^{Y}_{jl} \\
= & \sum_{i=1}^{I}\sum_{j=1}^{J}\sum_{k=1}^{I}\sum_{l=1}^{J} (\pi_{ij}-\pi_{i+}\pi_{+j})(\pi_{kl} - \pi_{k+}\pi_{+l})d^{X}_{ik}d^{Y}_{jl}.
\end{align*}
Under one-hot encodings for both $X$ and $Y$, i.e., $d^{X}_{ik} = \sqrt{2}I\{i\neq k\}$ and $d^{Y}_{jl} = \sqrt{2}I\{j\neq l\}$, we have 
\begin{align*}
\mathrm{dCov^{2}(X, Y)} = & 2\sum_{i=1}^{I}\sum_{j=1}^{J}\sum_{k\neq i}^{I}\sum_{l\neq j}^{J} (\pi_{ij}-\pi_{i+}\pi_{+j})(\pi_{kl} - \pi_{k+}\pi_{+l}) \\
= & 2\sum_{i=1}^{I}\sum_{j=1}^{J}(\pi_{ij}-\pi_{i+}\pi_{+j}) \left[ \sum_{k\neq i}^{I}\sum_{l\neq j}^{J}(\pi_{kl} - \pi_{k+}\pi_{+l}) \right],
\end{align*}
where 
\begin{align*}
\sum_{k\neq i}^{I}\sum_{l\neq j}^{J}(\pi_{kl} - \pi_{k+}\pi_{+l}) & = \sum_{k=1}^{I}\sum_{l=1}^{J}(\pi_{kl} - \pi_{k+}\pi_{+l}) - \sum_{l=1}^{J}(\pi_{il} - \pi_{i+}\pi_{+l}) - \sum_{k=1}^{I}(\pi_{kj} - \pi_{k+}\pi_{+j}) + (\pi_{ij}-\pi_{i+}\pi_{+j})  \\
& = \pi_{ij}-\pi_{i+}\pi_{+j}.
\end{align*}
Therefore, we have $dCov^{2}(X, Y) = 2\sum_{i=1}^{I}\sum_{j=1}^{J}(\pi_{ij}-\pi_{i+}\pi_{+j})^{2}$ under one-hot encodings.

\subsection{Proof of Lemma \ref{dvar}}\label{A2}
Using the following results
\begin{align*}
E(\|X_{1} - X_{2}\|) & = \sum_{i=1}^{I}\sum_{k=1}^{I}\pi_{i+}\pi_{k+}d^{X}_{ik} , \\
E(\|X_{1} - X_{2}\|^{2}) & = \sum_{i=1}^{I}\sum_{k=1}^{I}\pi_{i+}\pi_{k+}\left(d^{X}_{ik}\right)^{2}, \\
E(\|X_{1} - X_{2}\|\|X_{1} - X_{3}\|) & =  \sum_{i=1}^{I}\sum_{k=1}^{I}\sum_{m=1}^{I} \pi_{i+}\pi_{k+}\pi_{m+} d^{X}_{ik}d^{X}_{im},
\end{align*}
we have 
\begin{align*}
\mathrm{dVar}^{2}(X) = & \sum_{i=1}^{I}\sum_{k=1}^{I}\sum_{m=1}^{I}\sum_{q=1}^{I}\pi_{i+}\pi_{k+}\pi_{m+}\pi_{q+}\left(d^{X}_{ik}\right)^{2} + \sum_{i=1}^{I}\sum_{k=1}^{I}\sum_{m=1}^{I}\sum_{q=1}^{I}\pi_{i+}\pi_{k+}\pi_{m+}\pi_{q+}d^{X}_{ik}d^{X}_{mq} \\
& - 2 \sum_{i=1}^{I}\sum_{k=1}^{I}\sum_{m=1}^{I}\sum_{q=1}^{I}\pi_{i+}\pi_{k+}\pi_{m+}\pi_{q+}d^{X}_{ik}d^{X}_{im} \\
= & \sum_{i=1}^{I}\sum_{k=1}^{I} \pi_{i+}\pi_{k+} \sum_{m=1}^{I}\sum_{q=1}^{I}\pi_{m+}\pi_{q+}(d^{X}_{ik}-d^{X}_{im})(d^{X}_{ik}-d^{X}_{kq}) \\
= & \sum_{i=1}^{I}\sum_{k=1}^{I} \pi_{i+}\pi_{k+} (d^{X}_{ik} - \overline{d^{X}_{i}})(d^{X}_{ik} - \overline{d^{X}_{k}}),
\end{align*}
where 
\begin{equation*}
\overline{d^{X}_{i}} = \sum_{k=1}^{I}\pi_{k+}d^{X}_{ik}.
\end{equation*}
Note that under one-hot encodings, we have
\begin{align*}
\overline{d^{X}_{i}} & = \sqrt{2}(1-\pi_{i+}), \\
d^{X}_{ik} - \overline{d^{X}_{i}} & = \sqrt{2} \pi_{i+},  
\end{align*}
for $i\neq k$. Therefore the squared distance variance becomes
\begin{align*}
\mathrm{dVar}^{2}(X) & = 2\sum_{i=1}^{I}\sum_{k\neq i}^{I} \pi^{2}_{i+}\pi^{2}_{k+} + 2\sum_{i=1}^{I}\pi^{2}_{i+}(1-\pi_{i+})^{2} \\
& = 2\left( \sum_{i=1}^{I}\pi^{2}_{i+} \right)^{2} - 4\sum_{i=1}^{I}\pi^{3}_{i+} + 2\sum_{i=1}^{I}\pi^{2}_{i+}.
\end{align*}

\subsection{Proof of Lemma \ref{diff}}\label{A3}
Recall that 
\begin{align*}
\widehat{\mathrm{dCov}^{2}}(X, Y) & = \frac{T_{1}}{n^{2}} -\frac{2T_{2}}{n^{3}}+\frac{T_{3}}{n^{4}},\\
\widetilde{\mathrm{dCov}^{2}}(X, Y) & = \frac{T_{1}}{n(n-3)} -\frac{2T_{2}}{n(n-2)(n-3)}+\frac{T_{3}}{n(n-1)(n-2)(n-3)}, 
\end{align*}
where
\begin{align*}
T_{1} & = \sum_{i=1}^{I}\sum_{j=1}^{J}\sum_{k=1}^{I}\sum_{l=1}^{J} n_{ij}n_{kl}d^{X}_{ik}d^{Y}_{jl} , \\
T_{2} & =  \sum_{i=1}^{I}\sum_{j=1}^{J}\sum_{k=1}^{I}\sum_{l=1}^{J} n_{ij}n_{k+}n_{+l}d^{X}_{ik}d^{Y}_{jl}, \\
T_{3} & = \sum_{i=1}^{I}\sum_{j=1}^{J}\sum_{k=1}^{I}\sum_{l=1}^{J} n_{i+}n_{+j}n_{k+}n_{+l}d^{X}_{ik}d^{Y}_{jl}. \\
\end{align*}
By SLLN, it is straightforward to show
\begin{align*}
n\left[ \frac{T_{1}}{n^{2}} - \frac{T_{1}}{n(n-3)} \right] & \xrightarrow[]{a.s.} -3 \sum_{i=1}^{I}\sum_{j=1}^{J}\sum_{k=1}^{I}\sum_{l=1}^{J} \pi_{ij}\pi_{kl}d^{X}_{ik}d^{Y}_{jl}, \\
2n\left[ \frac{T_{2}}{n(n-2)(n-3)} - \frac{T_{2}}{n^{3}} \right] & \xrightarrow[]{a.s.} 10\sum_{i=1}^{I}\sum_{j=1}^{J}\sum_{k=1}^{I}\sum_{l=1}^{J}\pi_{ij}\pi_{k+}\pi_{+l}d^{X}_{ik}d^{Y}_{jl}, \\
n\left[ \frac{T_{3}}{n^{4}} - \frac{T_{3}}{n(n-1)(n-2)(n-3)} \right] & \xrightarrow[]{a.s.} -6\sum_{i=1}^{I}\sum_{j=1}^{J}\sum_{k=1}^{I}\sum_{l=1}^{J} \pi_{i+}\pi_{+j}\pi_{k+}\pi_{+l} d^{X}_{ik}d^{Y}_{jl}.
\end{align*}

Summarizing the results above, we have
\begin{equation*}
n\left[ \widehat{\mathrm{dCov}^{2}}(X, Y) - \widetilde{\mathrm{dCov}^{2}}(X, Y) \right] \xrightarrow[]{a.s.} \sum_{i=1}^{I}\sum_{j=1}^{J}\sum_{k=1}^{I}\sum_{l=1}^{J} (10\pi_{ij}\pi_{k+}\pi_{+l}-3\pi_{ij}\pi_{kl}-6\pi_{i+}\pi_{+j}\pi_{k+}\pi_{+l}) d^{X}_{ik}d^{Y}_{jl}.
\end{equation*}

\subsection{Proof of Theorem \ref{null}}\label{A4}
Since $H_{n} = \mathbf{1}_{n} - \frac{1}{n}\mathbf{1}_{n\times n}$ is the centering matrix, it follows that $H_{n}D_{n}^{X}H_{n}$ shares the same eigenvalues as $D_{n}^{X}H_{n}$. By sorting the samples in terms of $X$, the sample distance matrix $D_{n}^{X}$ and $D_{n}^{X}H_{n}$ can be simplified as follows 
\begin{equation*}
D_{n}^{X} = \begin{bmatrix} 
    0\cdot\mathbf{1}_{(n_{1+})\times (n_{1+})} & d^{X}_{12}\cdot\mathbf{1}_{(n_{1+})\times (n_{2+})} & \dots & d^{X}_{1I}\cdot\mathbf{1}_{(n_{1+})\times (n_{I+})} \\
    d^{X}_{21}\cdot \mathbf{1}_{(n_{2+})\times (n_{1+})} & 0\cdot\mathbf{1}_{(n_{2+})\times (n_{2+})} & \dots & d^{X}_{2I}\cdot\mathbf{1}_{(n_{2+})\times (n_{I+})} \\
    \vdots & \vdots  & \ddots & \vdots \\
    d^{X}_{I1}\cdot\mathbf{1}_{(n_{I+})\times (n_{1+})} &  d^{X}_{I2}\cdot\mathbf{1}_{(n_{I+})\times (n_{2+})} & \dots & 0\cdot\mathbf{1}_{(n_{I+})\times (n_{I+})}  
    \end{bmatrix},
 \end{equation*}   
 \begin{equation*}
D_{n}^{X}H_{n} = \begin{bmatrix} 
   -\overline{d^{X}_{1, n}}\cdot\mathbf{1}_{(n_{1+})\times (n_{1+})} & (d^{X}_{12}-\overline{d^{X}_{1, n}})\cdot\mathbf{1}_{(n_{1+})\times (n_{2+})} & \dots & (d^{X}_{1I}-\overline{d^{X}_{1, n}})\cdot\mathbf{1}_{(n_{1+})\times (n_{I+})} \\
    (d^{X}_{21}-\overline{d^{X}_{2, n}})\cdot \mathbf{1}_{(n_{2+})\times (n_{1+})} & -\overline{d^{X}_{2, n}}\cdot\mathbf{1}_{(n_{2+})\times (n_{2+})} & \dots & (d^{X}_{2I}-\overline{d^{X}_{2, n}})\cdot\mathbf{1}_{(n_{2+})\times (n_{I+})} \\
    \vdots & \vdots  & \ddots & \vdots \\
    (d^{X}_{I1}-\overline{d^{X}_{I, n}})\cdot\mathbf{1}_{(n_{I+})\times (n_{1+})} &  (d^{X}_{I2}-\overline{d^{X}_{I, n}})\cdot\mathbf{1}_{(n_{I+})\times (n_{2+})} & \dots & -\overline{d^{X}_{I, n}}\cdot\mathbf{1}_{(n_{I+})\times (n_{I+})}  
    \end{bmatrix},
 \end{equation*} 
 where $\overline{d^{X}_{i, n}} = \sum_{k=1}^{I}d^{X}_{ik}\widehat{\pi}_{k+}$ is the sample estimate of $\overline{d^{X}_{i}}$.  
 
Because $D_{n}^{X}H_{n}$ has a constant block structure, it has the same eigenvalues as the following matrix
  \begin{equation*}
\begin{bmatrix} 
    -\overline{d^{X}_{1, n}}n_{1+} & (d^{X}_{12}-\overline{d^{X}_{1, n}})\sqrt{n_{1+}n_{2+}} & \dots & (d^{X}_{1I}-\overline{d^{X}_{1, n}})\sqrt{n_{1+}n_{I+}} \\
    (d^{X}_{21}-\overline{d^{X}_{2, n}}) \sqrt{n_{2+}n_{1+})} & -\overline{d^{X}_{2, n}}n_{2+} & \dots & (d^{X}_{2I}-\overline{d^{X}_{2, n}})\sqrt{n_{2+}n_{I+})} \\
    \vdots & \vdots  & \ddots & \vdots \\
    (d^{X}_{I1}-\overline{d^{X}_{I, n}})\sqrt{n_{I+}n_{1+}} &  (d^{X}_{I2}-\overline{d^{X}_{I, n}})\sqrt{n_{I+}n_{2+}} & \dots & -\overline{d^{X}_{I, n}}n_{I+}
    \end{bmatrix}.
 \end{equation*}  
 
 Therefore the limiting eigenvalues of $H_{n}D_{n}^{X}H_{n}/n$ are same as those of matrix $Q^{X}$
  \begin{equation*}
  Q^{X} = 
 \begin{bmatrix} 
    -\overline{d^{X}_{1}}\pi_{1+} & (d^{X}_{12}-\overline{d^{X}_{1}})\sqrt{\pi_{1+}\pi_{2+}} & \dots & (d^{X}_{1I}-\overline{d^{X}_{1}})\sqrt{\pi_{1+}\pi_{I+}} \\
   (d^{X}_{21}-\overline{d^{X}_{2}})\sqrt{\pi_{2+}\pi_{1+}} &  -\overline{d^{X}_{2}}\pi_{2+} & \dots & (d^{X}_{2I}-\overline{d^{X}_{2}})\sqrt{\pi_{2+}\pi_{I+}} \\
    \vdots & \vdots  & \ddots & \vdots \\
    (d^{X}_{I1}-\overline{d^{X}_{I}})\sqrt{\pi_{I+}\pi_{1+}} &  (d^{X}_{I2}-\overline{d^{X}_{I}})\sqrt{\pi_{I+}\pi_{2+}} & \dots &  -\overline{d^{X}_{I}}\pi_{I+} 
    \end{bmatrix},
 \end{equation*}  
    
As shown by\linkedyearcite{}{Shen22}{Shen et al. (2022)}, for data with $I$ distinct values, $H_{n}D_{n}^{X}H_{n}/n$ is semi-positive definite with rank $I-1$. Let $\lambda_{1}\geq ... \geq \lambda_{I}$ be the eigenvalues of $Q^{X}$, we have $\lambda_{I}=0$ and the positive limiting eigenvalues of $H_{n}D_{n}^{X}H_{n}/n$ are $\{\lambda_{1}, ..., \lambda_{I-1}\}$, and 
\begin{equation*}
n\widetilde{\mathrm{dCov}^{2}}(X, Y) \xrightarrow{d} \sum_{i=1}^{I-1}\sum_{j=1}^{J-1} \lambda_{i}\mu_{j}(Z^{2}_{ij}-1),
\end{equation*}
where $Z^{2}_{ij}$ are $i.i.d.$ chi-squared random variables with $df=1$. 

For categorical distance variance, we have 
\begin{align*}
\mathrm{dVar}^{2}(X) & = \sum_{i=1}^{I}\sum_{k=1}^{I}\pi_{i+}\pi_{k+}(d^{X}_{ik}-\overline{d^{X}_{i}})(d^{X}_{ik}-\overline{d^{X}_{k}}) = \mathrm{trace}\left[ \left(Q^{X}\right)^{T}Q^{X} \right]  = \sum_{i=1}^{I-1}\lambda_{i}^{2}. \\
\mathrm{dVar}^{2}(Y) & = \sum_{j=1}^{J-1}\mu_{j}^{2}.
\end{align*}

Finally, by Remark \ref{a.s.} and Slutsky's theorem, Theorem \ref{null} is proved.

\subsection{Proof of Theorem \ref{alter}}\label{A5}
Our proof is based on multivariate delta method. First, by multivariate central limit theorem, we have 
\begin{equation} \label{mclt}
\sqrt{n}\left[ \mathrm{vec}(\widehat{\pi}) - \mathrm{vec}(\pi) \right] \xrightarrow{d} N(\mathbf{0},~ \Sigma),
\end{equation}
where $\Sigma$ is defined in \ref{sec23}. Next, we derive $D'_{ij}$, which is the first-order partial derivative of the squared distance covariance with respect to $\pi_{ij}$.  Recall that 
\begin{equation*}
\mathrm{dCov}^{2}(X, Y) = \sum_{q=1}^{I}\sum_{m=1}^{J}\sum_{k=1}^{I}\sum_{l=1}^{J}(\pi_{qm}-\pi_{q+}\pi_{+m})(\pi_{kl}-\pi_{k+}\pi_{+l})d^{X}_{qk}d^{Y}_{ml}.
\end{equation*}
The partial derivative $D'_{ij}$ can be partitioned into the following parts
\begin{align*}
D'_{ij, 1}: ~~& q = i, m = j, \\
D'_{ij, 2}: ~~& q = i, m \neq j, \\
D'_{ij, 3}: ~~& q \neq i, m = j.
\end{align*}
By simple calculations, we have
\begin{align*}
D'_{ij, 1}:  & = \sum_{k=1}^{I}\sum_{l=1}^{J}(\pi_{kl}-\pi_{k+}\pi_{+l})(1-\pi_{i+}-\pi_{+j})d^{X}_{ik}d^{Y}_{jl},\\
D'_{ij, 2}:  & = \sum_{k=1}^{I}\sum_{l=1}^{J}(\pi_{kl}-\pi_{k+}\pi_{+l}) \left( -\sum_{m\neq j}\pi_{+m}\right) d^{X}_{ik}d^{Y}_{ml}, \\
D'_{ij, 3}:  & = \sum_{k=1}^{I}\sum_{l=1}^{J}(\pi_{kl}-\pi_{k+}\pi_{+l}) \left( -\sum_{q\neq i}\pi_{q+}\right) d^{X}_{kq}d^{Y}_{jl}
\end{align*}
Summarizing the results above, we have
\begin{align*}
D'_{ij} & = D'_{ij, 1} + D'_{ij, 2} + D'_{ij, 3} \\
& = \sum_{k=1}^{I}\sum_{l=1}^{J}(\pi_{kl}-\pi_{k+}\pi_{+l})\left( d^{X}_{ik}d^{Y}_{jl}- \sum_{m=1}^{J}\pi_{+m}d^{X}_{ik}d^{Y}_{ml} - \sum_{q=1}^{I}\pi_{q+}d^{X}_{kq}d^{Y}_{jl} \right) \\
& = \sum_{k=1}^{I}\sum_{l=1}^{J} (\pi_{kl} - \pi_{k+}\pi_{+l}) \left(d^{X}_{ik}d^{Y}_{jl} - d^{X}_{ik}\overline{d^{Y}_{l}} - d^{Y}_{jl}\overline{d^{X}_{k}} \right).
\end{align*}
By multivariate delta method and Equation \ref{mclt}, we have
\begin{equation*}
\sqrt{n}\widehat{\mathrm{dCov}^{2}}(X, Y) \xrightarrow{d} N\left\{ \mathrm{dCov}^{2}(X, Y), ~ [ \mathrm{vec}(D') ]^{\intercal} \Sigma \mathrm{vec}(D') \right\}.
\end{equation*}
By Lemma \ref{diff}, we have $\sqrt{n} [ \widehat{dCov^{2}}(X, Y) - \widetilde{dCov^{2}}(X, Y) ]  \xrightarrow[]{a.s.} 0$, therefore
\begin{equation*}
\sqrt{n}\widetilde{\mathrm{dCov}^{2}}(X, Y) \xrightarrow{d} N\left\{ \mathrm{dCov}^{2}(X, Y), ~ [ \mathrm{vec}(D') ]^{\intercal} \Sigma \mathrm{vec}(D') \right\}.
\end{equation*}
Finally, by Remark \ref{a.s.} and Slutsky's theorem, Theorem \ref{alter} is proved.

\subsection{Proof of Lemma \ref{concentrate}}\label{A6}
First, we bound $\partial dCor^{2}(X^{s}, Y)/\partial \pi_{i_{s}j}$, where $X^{s}\in \mathcal{S}_{T}$. By quotient rule, we have
\begin{equation*}
\frac{\partial \mathrm{dCor}^{2}(X^{s}, Y)}{\partial \pi_{i_{s}j}} = \frac{\frac{\partial\mathrm{dCov}^{2}(X^{s}, Y)}{\partial \pi_{i_{s}j}} \mathrm{dVar}(X^{s}) \mathrm{dVar}(Y) - \frac{\partial\mathrm{dVar}(Y)}{\partial \pi_{i_{s}j}}\mathrm{dCov}^{2}(X^{s}, Y)\mathrm{dVar}(X^{s})- \frac{\partial\mathrm{dVar}(X^{s})}{\partial \pi_{i_{s}j}}\mathrm{dCov}^{2}(X^{s}, Y)\mathrm{dVar}(Y)}{\mathrm{dVar}^{2}(X^{s}) \mathrm{dVar}^{2}(Y)}.
\end{equation*}
Under rescaled distances, i.e., $\max_{i_{s},k_{s}}d^{X^{s}}_{i_{s}k_{s}} = \max_{j,l}d^{Y}_{jl} = 1$, and Condition (C2) we have 
\begin{align*}
& \delta^{2}_{min}\leq \left| \mathrm{dCov}^{2}(X^{s}, Y) \right|  \leq 1, \\
& \sigma^{2}_{min}\leq \left| \mathrm{dVar}(X^{s}) \right|  \leq 1, \\
 & \sigma^{2}_{min}\leq\left| \mathrm{dVar}(Y) \right|  \leq 1.
\end{align*}
Using Theorem \ref{alter}, we have
\begin{equation*}
\frac{\partial \mathrm{dCov}^{2}(X^{s}, Y)}{\partial \pi_{i_{s}j}} =  \sum_{k_{s}=1}^{I_{s}}\sum_{l=1}^{J} (\pi_{k_{s}l} - \pi_{k_{s}+}\pi_{+l}) \left(d^{X^{s}}_{i_{s}k_{s}}d^{Y}_{jl} - d^{X^{s}}_{i_{s}k_{s}}\overline{d^{Y}_{l}} - d^{Y}_{jl}\overline{d^{X^{s}}_{k_{s}}} \right),
\end{equation*}
and 
\begin{equation*}
\left| \frac{\partial \mathrm{dCov}^{2}(X^{s}, Y)}{\partial \pi_{i_{s}j}} \right| \leq 2\sum_{k_{s}=1}^{I_{s}}\sum_{l=1}^{J} (\pi_{k_{s}l} + \pi_{k_{s}+}\pi_{+l}) = 4.
\end{equation*}
Similarly, we have
\begin{equation*}
\frac{\partial \mathrm{dVar}^{2}(X^{s})}{\partial \pi_{i_{s}j}} = 2 \sum_{k_{s}=1}^{I_{s}}\pi_{k_{s}+}\left( d^{X^{s}}_{i_{s}k_{s}}\overline{d^{X^{s}}_{i_{s}}} \right)\left( d^{X^{s}}_{i_{s}k_{s}}\overline{d^{X^{s}}_{k_{s}}} \right),
\end{equation*}
and 
\begin{align*}
& \left| \frac{\partial \mathrm{dVar}^{2}(X^{s})}{\partial \pi_{i_{s}j}}  \right| \leq 2, \\
& \left| \frac{\partial \mathrm{dVar}^{2}(Y)}{\partial \pi_{i_{s}j}}  \right|  \leq 2, \\
& \left| \frac{\partial \mathrm{dVar}(X^{s})}{\partial \pi_{i_{s}j}}  \right|  \leq \frac{1}{\sigma^{2}_{min}}, \\
& \left| \frac{\partial \mathrm{dVar}(Y)}{\partial \pi_{i_{s}j}}  \right|  \leq \frac{1}{\sigma^{2}_{min}}.
\end{align*}
Summarizing the results above, we have 
\begin{align*}
\left| \frac{\partial \mathrm{dCor}^{2}(X^{s}, Y)}{\partial \pi_{i_{s}j}} \right| & \leq \frac{1}{\sigma^{8}_{min}}\left( 4+\frac{1}{\sigma^{2}_{min}} + \frac{1}{\sigma^{2}_{min}} \right) \\
& = \frac{4}{\sigma^{8}_{min}} + \frac{2}{\sigma^{10}_{min}}.
\end{align*}
As $dCor^{2}(X^{s}, Y)$ is a continuous and differentiable function of $\{\pi_{i_{s}j}\}_{1\leq i_{s}\leq I_{s}, 1\leq j\leq J}$, we have 
\begin{align*}
\max_{s}\left | \widehat{\mathrm{dCor}^{2}}(X^{s}, Y) - \mathrm{dCor}^{2}(X^{s}, Y) \right | & \leq \left(  \frac{4}{\sigma^{8}_{min}} + \frac{2}{\sigma^{10}_{min}} \right) \max_{s}\sum_{i_{s}=1}^{I_{s}}\sum_{j=1}^{J} \left | \widehat{\pi}_{i_{s}j} - \pi_{i_{s}j} \right| \\
& \leq I_{max}J \left(  \frac{4}{\sigma^{8}_{min}} + \frac{2}{\sigma^{10}_{min}} \right) \max_{j}\max_{s}\max_{i_{s}}\left | \widehat{\pi}_{i_{s}j} - \pi_{i_{s}j} \right|.
\end{align*}
It suffices to prove the uniform consistency of $\widehat{\pi}_{i_{s}j}$. Similar to\linkedyearcite{}{Huangetal}{Huang et al. (2014)} and\linkedyearcite{}{Zhang2025}{Zhang (2025)}, let
\begin{equation*}
Z_{mi_{s}j} = I\{X^{s}_{m} = i_{s}\} - \pi_{i_{s}j},
\end{equation*}
where $m=1,...,n$ is the sample index. We have 
\begin{equation*}
\widehat{\pi}_{i_{s}j} - \pi_{i_{s}j} = \frac{1}{n}\sum_{m=1}^{n}Z_{mi_{s}j}.
\end{equation*}
By Bernstein's inequality, for any $\epsilon>0$, we have 
\begin{equation*}
P\left( \left| \frac{1}{n}\sum_{m=1}^{n} Z_{mi_{s}j} \right|>\epsilon \right) \leq 2\exp\left( -\frac{6n\epsilon^2}{4\epsilon + 3} \right).
\end{equation*} 
Note that $\widehat{\pi}_{i_{s}j} - \pi_{i_{s}j} = (1/n)\sum_{m=1}^{n} Z_{mi_{s}j} $, we have
\begin{align*}
P\left( \max_{j}\max_{s}\max_{i_{s}}\left| \widehat{\pi}_{i_{s}j} - \pi_{i_{s}j} \right|>\epsilon \right) & \leq \sum_{j=1}^{J}\sum_{s=1}^{S}\sum_{i_{k}=1}^{I_{k}} P\left( \frac{1}{n}\left| \sum_{m=1}^{n} Z_{mi_{k}j} \right| > \epsilon \right)  \\
& \leq  2S I_{max} J\cdot\exp\left( -\frac{6n\epsilon^2}{4\epsilon + 3} \right) \\
& \leq 2I_{max} J\cdot\exp\left( \log S-\frac{6n\epsilon^2}{4\epsilon + 3} \right). 
\end{align*}
By Condition (C3), $\widehat{\pi}_{i_{s}j}$ is uniformly consistent, therefore $\widehat{dCor^{2}}(X^{s}, Y)$ is uniformly consistent. To be specific, we have the following probability inequality
\begin{align*}
P\left(\max_{s} \left | \widehat{\mathrm{dCor}^{2}}(X^{s}, Y) - \mathrm{dCor}^{2}(X^{s}, Y) \right | >\epsilon \right) & \leq P\left( \max_{j}\max_{s}\max_{i_{s}}\left| \widehat{\pi}_{i_{s}j} - \pi_{i_{s}j} \right| > \frac{\epsilon}{I_{max}J \left(  \frac{4}{\sigma^{8}_{min}} + \frac{2}{\sigma^{10}_{min}} \right)} \right) \\
& \leq 2I_{max} J\cdot\exp\left[ \log S-\frac{6n\epsilon^2}{4\epsilon I_{max}J \left(  \frac{4}{\sigma^{8}_{min}} + \frac{2}{\sigma^{10}_{min}} \right) + 3I^{2}_{max}J^{2} \left(  \frac{4}{\sigma^{8}_{min}} + \frac{2}{\sigma^{10}_{min}} \right)^{2}} \right].
\end{align*}

\subsection{Proof of Theorem \ref{consistent}}\label{A7}
By Condition (C2), there exists a positive constant, $\delta^{2}_{min}>0$, such that $\min_{X^{s}\in \mathcal{S}_{T}} dCov^{2}(X^{s}, Y)>\delta^{2}_{min}$. As $dVar^{2}(X^{s}) \leq 1$ and $dVar^{2}(Y) \leq 1$ under rescaled distances, we have $\min_{X^{s}\in \mathcal{S}_{T}} dCor^{2}(X^{s}, Y) > \delta^{2}_{min}$. Let $C = \delta^{2}_{min}/2$, by the uniform consistency of $\widehat{dCor^{2}}(X^{s}, Y)$, we have $\mathcal{S}_{T}\subset\widehat{S}(C)$ with probability converging to 1. Otherwise, there exists $r$, such that $X^{r}\in \mathcal{S}_{T}$, but $X^{r} \not\in \widehat{S}(C)$, therefore we have $dCor^{2}(X^{r}, Y)>\delta^{2}_{min}$ and $\widehat{dCor^{2}}(X^{r}, Y) < \delta^{2}_{min}/2$, so $ | \widehat{dCor^{2}}(X^{r}, Y) - dCor^{2}(X^{r}, Y) | > \delta^{2}_{min}/2$. On the other hand, by Lemma \ref{concentrate} we know $\widehat{dCor^{2}}(X^{r}, Y)$ is uniformly consistent. Let $\epsilon =  \delta^{2}_{min}/2$, we have
\begin{equation*}
P\left( \left | \widehat{\mathrm{dCor}^{2}}(X^{r}, Y) - \mathrm{dCor}^{2}(X^{r}, Y) \right | > \delta^{2}_{min}/2 \right) \rightarrow 0,
\end{equation*} 
as $n\rightarrow \infty$, which leads to a contradiction.

Similarly, we have $\widehat{S}(C)\subset\mathcal{S}_{T}$. Otherwise, there exists $r$, such that $X^{r}\in \widehat{S}(C)$, but $X^{r} \not\in \mathcal{S}_{T}$, which means $dCor^{2}(X^{r}, Y) = 0$ but $\widehat{dCor^{2}}(X^{r}, Y) > \delta^{2}_{min}/2$. This causes the same contradiction with the uniform consistency by Lemma \ref{concentrate}. As a result, we know $P[\widehat{S}(C) = \mathcal{S}_{T}]\rightarrow 1$ with $C = \delta^{2}_{min}/2$, as $n\rightarrow \infty$. This completes the proof.

\section*{Conflict of Interest Statement}
\noindent
The author states that there is no conflict of interest.

\section*{Funding Declaration}
\noindent
The work was supported by an NSF DBI Biology Integration Institute (BII) grant (award no. 2119968).

\newpage
\section*{Figures and Tables}
\begin{figure}[!htbp]
\begin{center}
\includegraphics[scale=0.4]{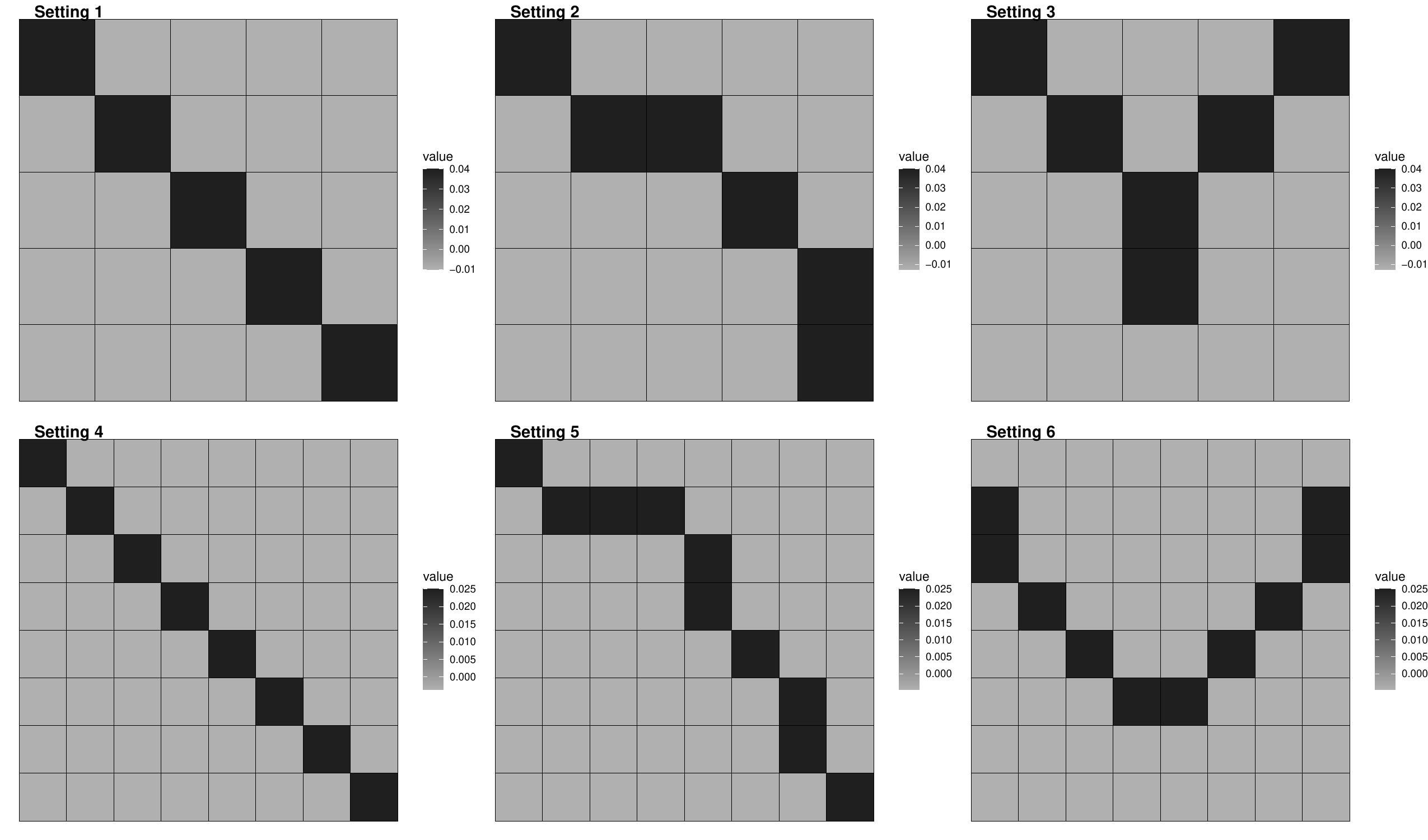}
\end{center}
\caption{Dependence patterns in the six simulation settings, with darker shades indicating a greater departure from independence.
}
\end{figure}

\newpage
\begin{figure}[!htbp]
\begin{center}
\includegraphics[scale=0.6]{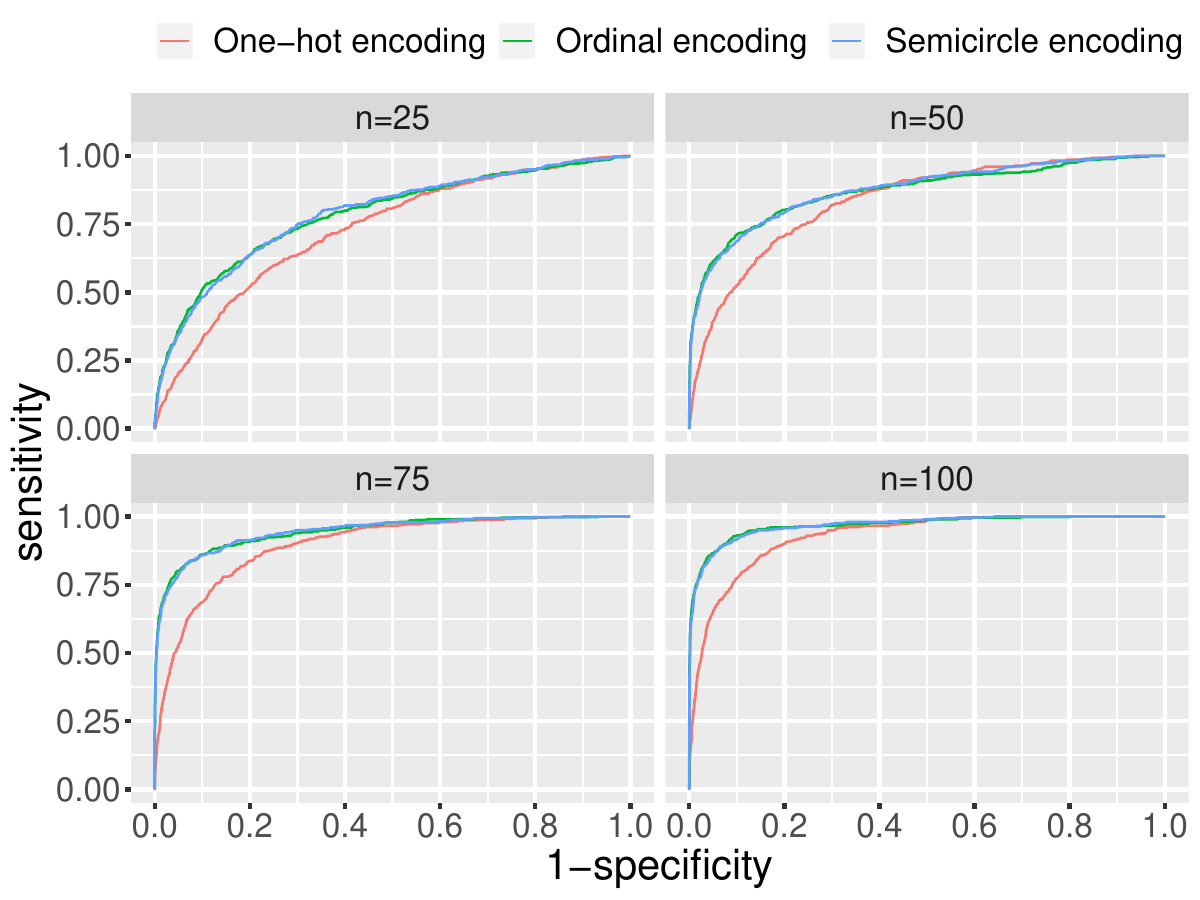}
\end{center}
\caption{ROC curves for distance correlation screening under three encodings (one-hot, ordinal, and semicircle) in Setting 1.
}
\end{figure}

\newpage
\begin{figure}[!htbp]
\begin{center}
\includegraphics[scale=0.6]{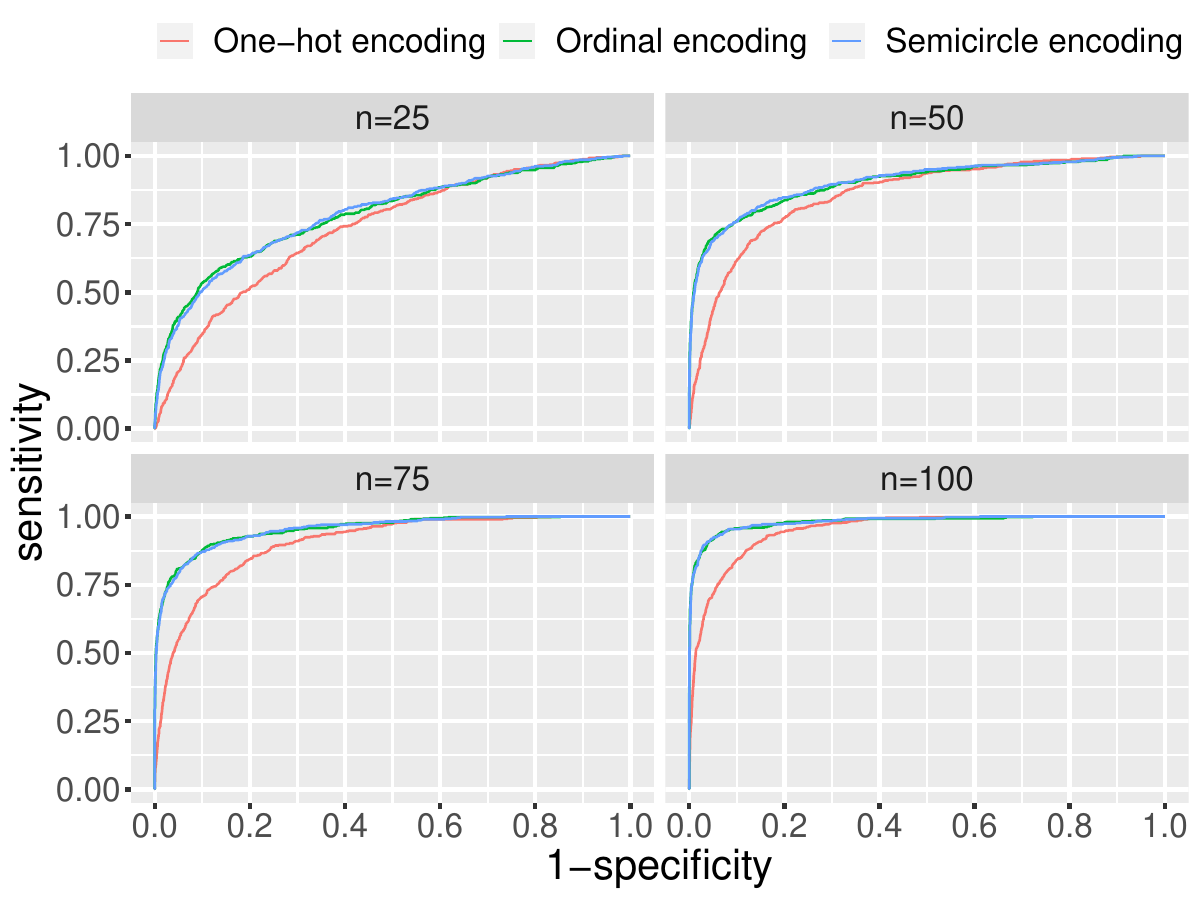}
\end{center}
\caption{ROC curves for distance correlation screening under three encodings (one-hot, ordinal, and semicircle) in Setting 2.
}
\end{figure}

\newpage
\begin{figure}[!htbp]
\begin{center}
\includegraphics[scale=0.6]{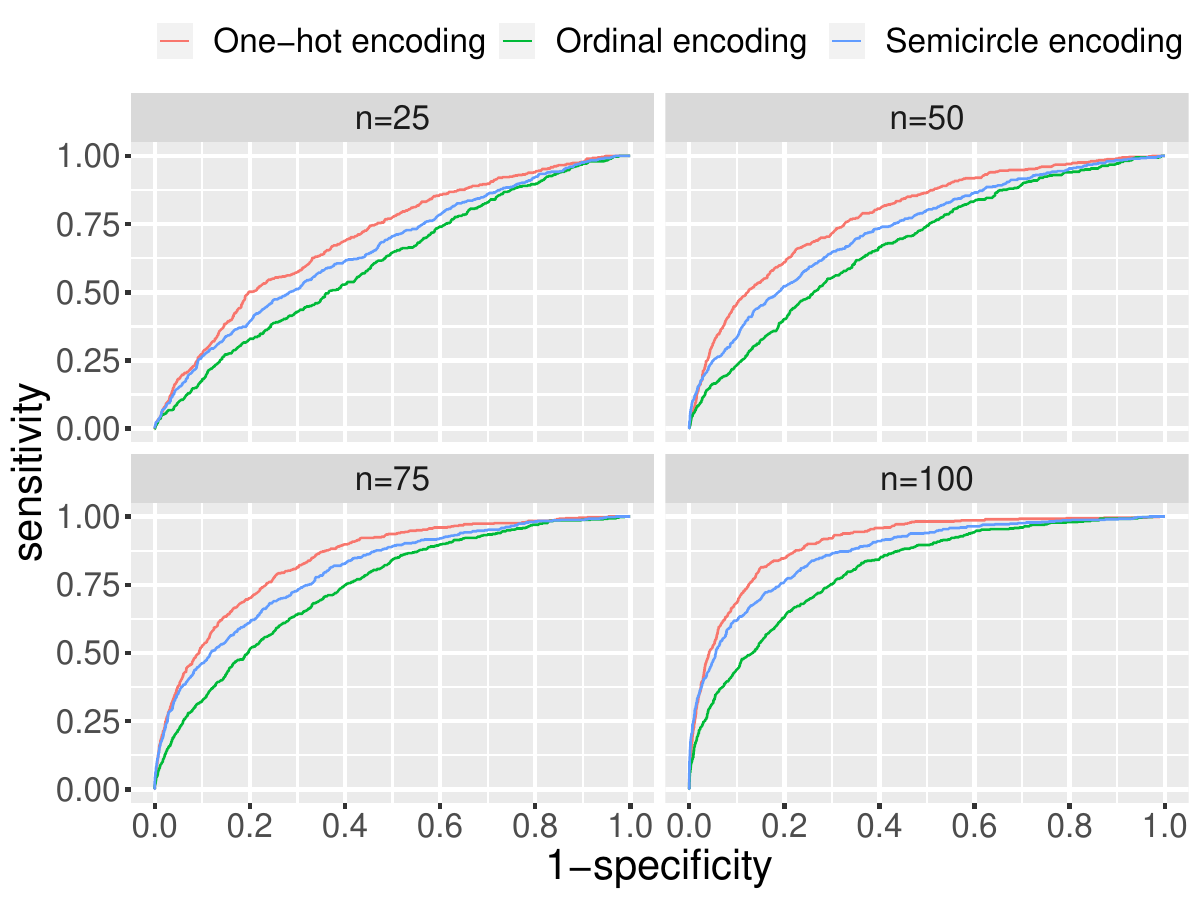}
\end{center}
\caption{ROC curves for distance correlation screening under three encodings (one-hot, ordinal, and semicircle) in Setting 3.
}
\end{figure}

\newpage
\begin{figure}[!htbp]
\begin{center}
\includegraphics[scale=0.6]{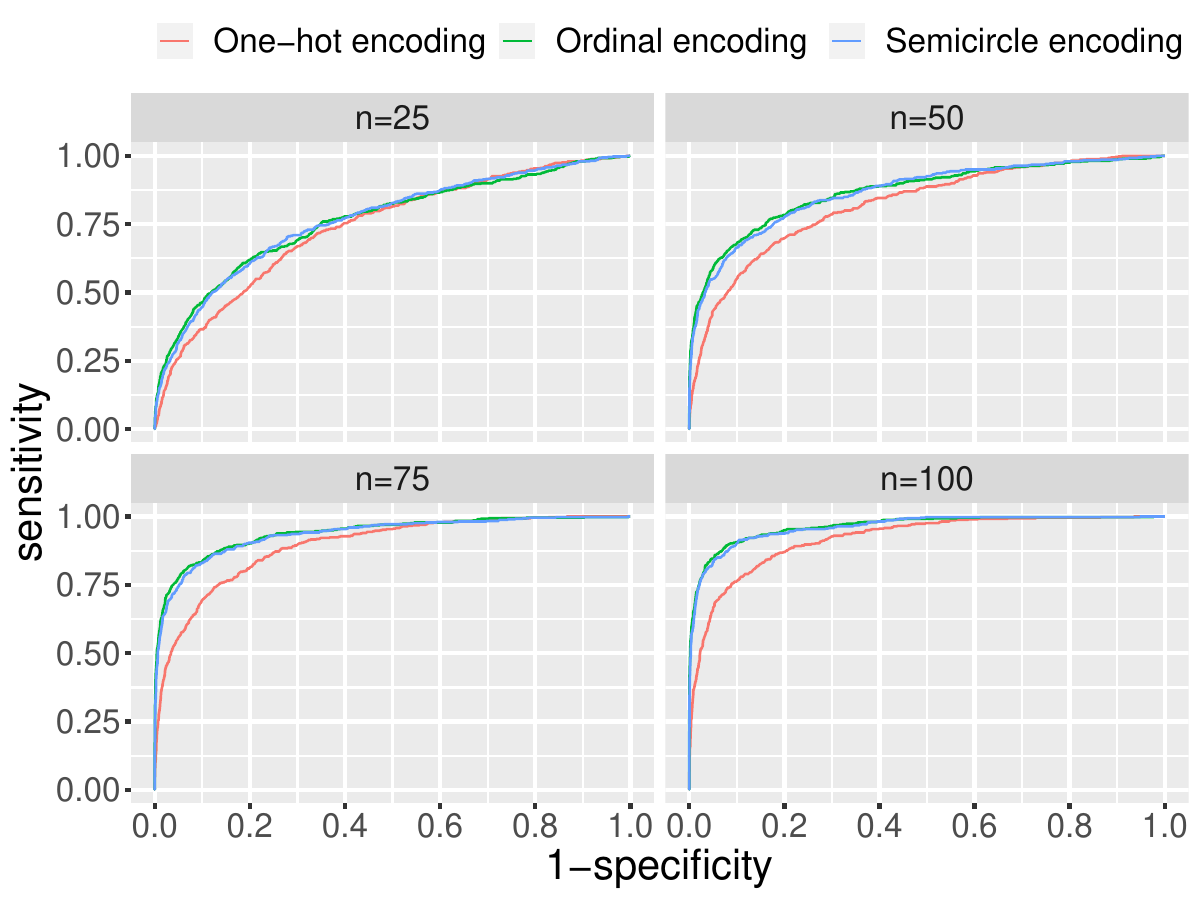}
\end{center}
\caption{ROC curves for distance correlation screening under three encodings (one-hot, ordinal, and semicircle) in Setting 4.
}
\end{figure}

\newpage
\begin{figure}[!htbp]
\begin{center}
\includegraphics[scale=0.6]{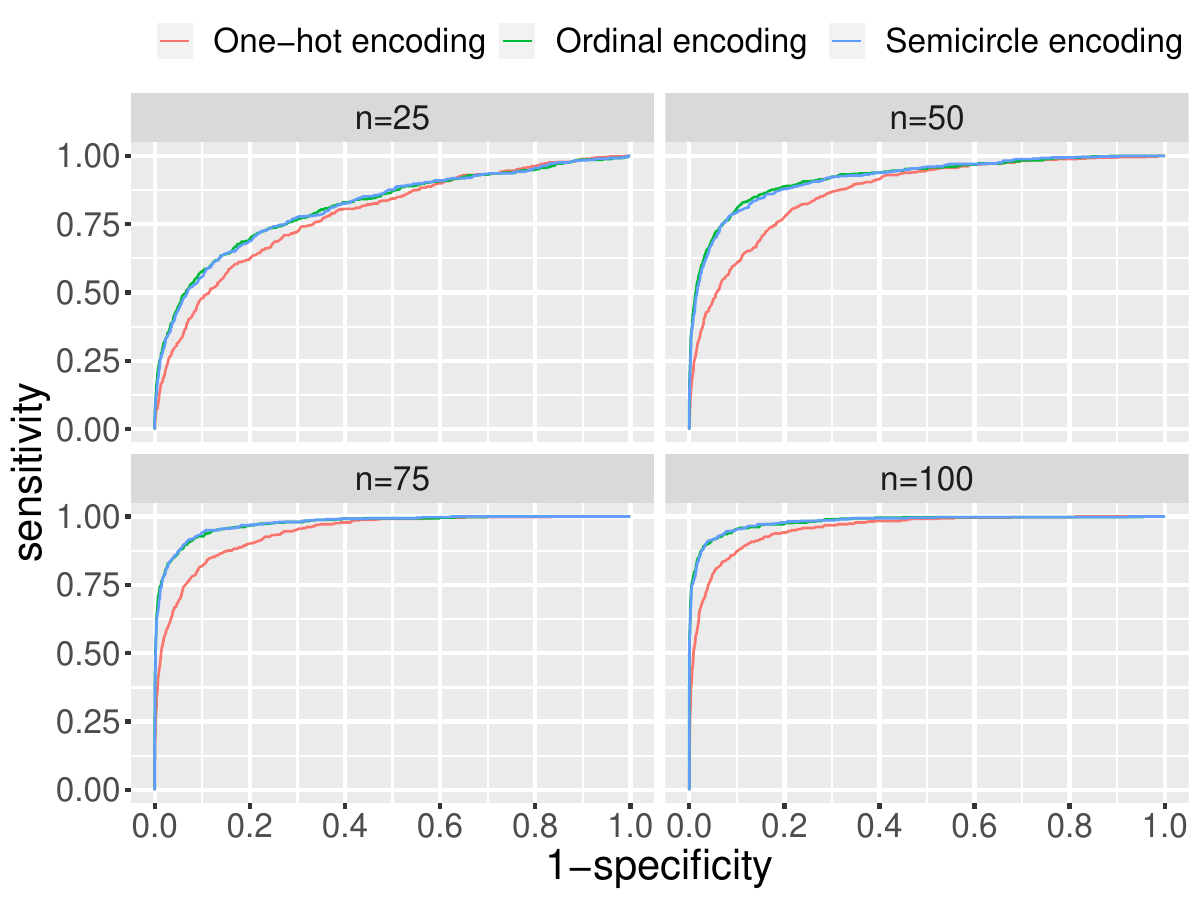}
\end{center}
\caption{ROC curves for distance correlation screening under three encodings (one-hot, ordinal, and semicircle) in Setting 5.
}
\end{figure}

\newpage
\begin{figure}[!htbp]
\begin{center}
\includegraphics[scale=0.6]{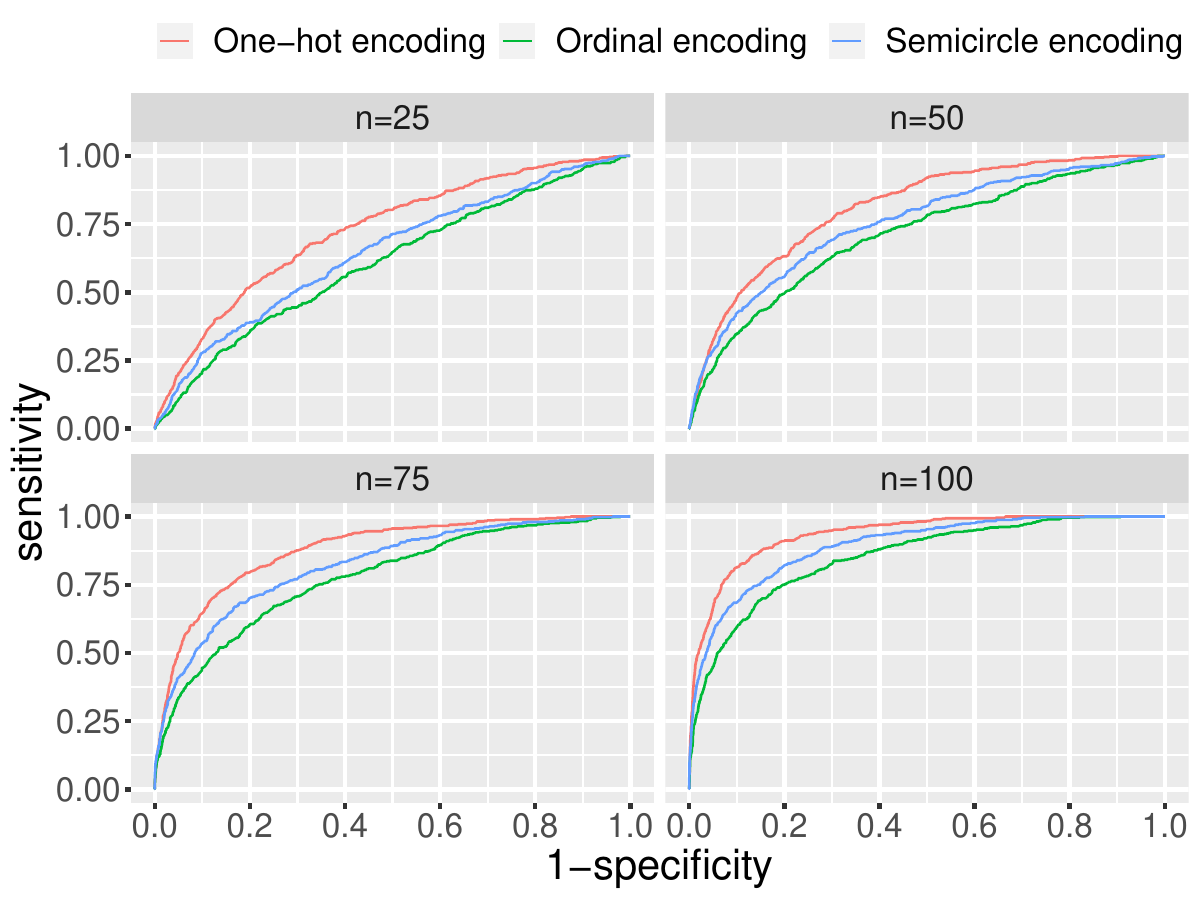}
\end{center}
\caption{ROC curves for distance correlation screening under three encodings (one-hot, ordinal, and semicircle) in Setting 6.
}
\end{figure}

\begin{sidewaystable}
\begin{table}[H]  \small
\centering 
 \rowcolors{1}{}{lightgray}
\begin{tabular}{p{12mm}|p{10mm}|p{23mm}|p{23mm}|p{25mm}|p{30mm}|p{30mm}|p{32mm}}
  \hline  \hline
Setting  & $n$ & AUC one-hot & AUC ordinal & AUC semicircle & (sens, spec) one-hot & (sens, spec) ordinal & (sens, spec) semicircle \vspace{2mm} \\ 
  \hline
1 &	25 &	0.731 & 0.785 & 0.787 & (0.286, 0.915) &  (0.424, 0.929) & (0.430, 0.930)\vspace{2mm} \\ 
1 & 50 & 0.835	& 0.872 & 0.875 & (0.490, 0.918) & (0.532, 0.975) & (0.536, 0.974)\vspace{2mm} \\ 
1 & 75 & 0.897 & 0.946 & 0.946 & (0.594, 0.937) & (0.673, 0.982) & (0.669, 0.983)\vspace{2mm} \\ 
1 & 100 & 0.928 & 0.969 & 0.969 & (0.638, 0.954) & (0.727, 0.985) & (0.725, 0.985)\vspace{2mm} \\  
\hline
2 & 25 & 0.734 & 0.785 & 0.787 & (0.286, 0.909) & (0.412, 0.930) & (0.418, 0.929)\vspace{2mm} \\  
2 & 50 & 0.859 & 0.903 & 0.906 & (0.438, 0.924) & (0.510, 0.971) & (0.519, 0.969)\vspace{2mm} \\  
2 & 75 & 0.902	& 0.953 & 0.953 & (0.626, 0.939) & (0.696, 0.984) & (0.708, 0.984) \vspace{2mm}\\  
2 & 100 & 0.951 & 0.981 & 0.982 & (0.730, 0.946) & (0.826, 0.986) & (0.830, 0.987) \vspace{2mm}\\  
  \hline
3 & 25 & 0.703 & 0.606 & 0.655 & (0.222, 0.922) & (0.142, 0.911) & (0.198, 0.913)\vspace{2mm} \\  
3 & 50 & 0.787 & 0.678 & 0.733 & (0.388, 0.924) & (0.198, 0.924) & (0.300, 0.931)\vspace{2mm} \\  
3 & 75 & 0.840 & 0.741 & 0.800 & (0.450, 0.931) & (0.308, 0.933) & (0.416, 0.937)\vspace{2mm} \\  
3 & 100 & 0.904 & 0.805 & 0.867 & (0.596, 0.937) & (0.346, 0.940) & (0.491, 0.960) \\
   \hline  \hline
\end{tabular}
\caption{AUC, sensitivity, and specificity of distance correlation screening under one-hot, ordinal, and semicircle encodings in Settings 1-3 ($I_{s}=J=5$). Sensitivity and specificity values are based on change-point selected tuning parameters.}
\end{table}
\end{sidewaystable}

\begin{sidewaystable}
\begin{table}[H]  \small
\centering 
 \rowcolors{1}{}{lightgray}
\begin{tabular}{p{12mm}|p{10mm}|p{23mm}|p{23mm}|p{25mm}|p{30mm}|p{30mm}|p{32mm}}
  \hline  \hline
Setting  & $n$ & AUC one-hot & AUC ordinal & AUC semicircle & (sens, spec) one-hot & (sens, spec) ordinal & (sens, spec) semicircle  \vspace{2mm}\\ 
  \hline
4 & 25 & 0.742 & 0.767 & 0.768 & (0.314, 0.930) & (0.397, 0.927) & (0.396, 0.924)\vspace{2mm} \\  
4 & 50 & 0.824 & 0.871 & 0.868 & (0.446, 0.939) & (0.534, 0.966) & (0.532, 0.965)\vspace{2mm} \\  
4 & 75 & 0.894 & 0.941 & 0.936 & (0.548, 0.953) & (0.622, 0.980) & (0.620, 0.979)\vspace{2mm} \\  
4 & 100 & 0.92 & 0.965 & 0.962 & (0.612, 0.959) & (0.732, 0.986) & (0.733, 0.983) \vspace{2mm} \\  
\hline
5 & 25 & 0.785 & 0.816 & 0.815 & (0.394, 0.912) & (0.492, 0.943) & (0.488, 0.945)\vspace{2mm} \\  
5 & 50 & 0.870 & 0.922 & 0.919 & (0.486, 0.945) & (0.572, 0.973) & (0.568, 0.974)\vspace{2mm} \\  
5 & 75 & 0.940 & 0.974 & 0.975 & (0.584, 0.966) & (0.743, 0.982) & (0.746, 0.980)\vspace{2mm} \\  
5 & 100 & 0.955 & 0.981 & 0.982 & (0.704, 0.975) & (0.841, 0.984) & (0.842, 0.984) \vspace{2mm} \\  
  \hline
6 & 25 & 0.721 & 0.61 & 0.651 & (0.252, 0.929) & (0.208, 0.907) & (0.244, 0.912) \vspace{2mm} \\  
6 & 50 & 0.812 & 0.713 & 0.756 & (0.388, 0.927) & (0.306, 0.915) & (0.354, 0.924) \vspace{2mm} \\  
6 & 75 & 0.878 & 0.777 & 0.823 & (0.516, 0.937) & (0.392, 0.924) & (0.466, 0.939) \vspace{2mm} \\  
6 & 100 & 0.934 & 0.852 & 0.894 & (0.638, 0.954) & (0.516, 0.937) & (0.575, 0.955) \\  
   \hline  \hline
\end{tabular}
\caption{AUC, sensitivity, and specificity of distance correlation screening under one-hot, ordinal, and semicircle encodings in Settings 4-6 ($I_{s}=J=8$). Sensitivity and specificity values are based on change-point selected tuning parameters.}
\end{table}
\end{sidewaystable}

\newpage
\begin{table}[H]  \small
\centering 
 \rowcolors{1}{}{lightgray}
\begin{tabular}{p{24mm}|p{15mm}|p{110mm}}
  \hline  \hline
Variable & Type & Survey question  \vspace{2mm}\\ 
  \hline
CANTRUST & ordinal & People can be trusted or cannot be too careful  \vspace{1mm}\\  
CODEG & ordinal & Partner's highest degree  \vspace{1mm}\\ 
COLSCI & nominal & Respondent has taken any college-level science course  \vspace{1mm}\\ 
CONBIZ & ordinal & Confidence in business and industry \vspace{1mm}\\
CRACK30 & ordinal & Respondent last use crack cocaine\vspace{1mm} \\
DEGREE & ordinal & Respondent's highest degree \vspace{1mm}\\
DWELOWN & nominal & Respondent owns or rents home \vspace{1mm}\\
ENDSMEET & ordinal & How difficult is it for respondent's household to make ends meet \vspace{1mm}\\
FAIR & ordinal & People fair or try to take advantage\vspace{1mm}\\
FINRELA & ordinal & Opinion of family income\vspace{1mm} \\
FUCITZN & nominal & Is respondent planning/applying for US citizenship or not\vspace{1mm} \\
GOODLIFE & ordinal & Standard of living of respondent will improve\vspace{1mm} \\
HANDMOVE & nominal & Respondent performs forceful hand movements \vspace{1mm}\\
HEALTHISSP & ordinal & Respondent's health in general \vspace{1mm}\\
HLTHPHYS & ordinal & Respondent's physical health \vspace{1mm}\\
HVYLIFT & nominal & Respondent does repeated lifting \vspace{1mm}\\
INCUSPOP & ordinal & Estimated income status of housing unit \vspace{1mm}\\
KNWEXEC & ordinal & Does respondent know someone who is a executive at a large company \vspace{1mm}\\
KNWLAWYR & ordinal & Does respondent know someone who is a lawyer \vspace{1mm}\\
KNWMW4 & nominal & Gender identity of fourth person \vspace{1mm} \\
MAR2 & nominal & Marital status of second person\vspace{1mm} \\
NEISAFE & ordinal & How safe respondent thinks neighborhood is\vspace{1mm} \\
PADEG & ordinal & Father's highest degree\vspace{1mm} \\
PARTLSC & ordinal & Respondent has participated in organizations for sport or culture\vspace{1mm} \\
PHYEFFRT & ordinal & Rate physical effort \vspace{1mm}\\
QUALLIFE & ordinal & Respondent's quality of life\vspace{1mm}\\
RELATE2 & nominal & Relationship of second person to household head \vspace{1mm}\\
RELHHD6 & nominal & Relationship of 6th person to household head\vspace{1mm} \\
SATFIN & ordinal & Satisfaction with financial situation\vspace{1mm} \\
SPDEG & ordinal & Spouse's highest degree  \vspace{1mm}\\ 
USCITZN & nominal & Is respondent US citizen\vspace{1mm} \\
WRKHOME & ordinal & How often respondent works at home\vspace{1mm} \\
   \hline  \hline
\end{tabular}
\caption{Identified categorical variables related to subjective class identification.}
\end{table}

\end{document}